\documentclass[usenatbib]{mn2e}
\usepackage{graphicx}
\usepackage{graphics}
\usepackage{txfonts}
\usepackage{AMSmath}
\usepackage{url}
\newcommand{\kms} {\mbox{${\rm km\,s^{-1}}$}}

\title[Water deuterium fractionation 
in the high-mass star-forming region G34.26+0.15 
]{{Water deuterium fractionation 
in the high-mass star-forming region G34.26+0.15 
based on \textit{Herschel}/HIFI data}
}

\author[A. Coutens, C. Vastel, U. Hincelin et al.]{A. Coutens$^{1,2}$\thanks{E-mail:
acoutens@nbi.dk}, C. Vastel$^{3,4}$, U. Hincelin$^{5}$, E. Herbst$^{5}$, D. C. Lis$^{6,7}$, L. Chavarr\'ia$^{8}$, 
  \newauthor M. G\'erin$^{9}$, F. F. S. van der Tak$^{10,11}$, C. M. Persson$^{12}$, P. F. Goldsmith$^{13}$, E. Caux$^{3,4}$ \\ 
$^{1}$ Niels Bohr Institute, University of Copenhagen, Juliane Maries Vej 30, DK-2100 Copenhagen \O, Denmark \\
$^{2}$ Centre for Star and Planet Formation, Natural History Museum of Denmark, University of Copenhagen, \O ster Voldgade 5-7, \\
DK-1350 Copenhagen K, Denmark \\
$^{3}$ Universit\'e de Toulouse, UPS-OMP, IRAP, Toulouse, France \\
$^{4}$ CNRS, Institut de Recherche en Astrophysique et Plan\'etologie, 9 Av. Colonel Roche, BP 44346, 31028 Toulouse Cedex 4, France \\
$^{5}$ Department of Chemistry, University of Virginia, McCormick Road, Charlottesville, VA 22904, USA \\
$^{6}$ California Institute of Technology, Cahill Center for Astronomy and Astrophysics 301-17, Pasadena, CA 91125, USA\\
$^{7}$ Sorbonne Universit\'{e}s, Universit\'{e} Pierre et Marie Curie, Paris 6, CNRS, Observatoire de Paris, UMR 8112, LERMA, Paris, France \\
$^{8}$ Universidad de Chile - CONICYT, Camino del Observatorio 1515, Las Condes, Santiago \\
$^{9}$ LERMA-LRA, UMR 8112 du CNRS, Observatoire de Paris, Ecole Normale Sup\'erieure, UPMC \& UCP, 24 rue Lhomond, 75231 Paris Cedex 05, France \\
$^{10}$ SRON Netherlands Institute for Space Research, Landleven 12, 9747 AD Groningen, The Netherlands \\
$^{11}$ Kapteyn Astronomical Institute, University of Groningen, 9700 AV Groningen, The Netherlands \\
$^{12}$ Chalmers University of Technology, Department of Earth and Space Sciences, Onsala Space Observatory, 43992 Onsala, Sweden \\
$^{13}$ Jet Propulsion Laboratory, California Institute of Technology, Pasadena, CA 91125, USA 
}
\begin{document}

\date{Accepted xxx. Received xxx; in original form xxx}


\maketitle

\label{firstpage}

\begin{abstract}
Understanding water deuterium fractionation is important for constraining the mechanisms of water formation in interstellar clouds.
Observations of HDO and H$_2^{18}$O transitions were carried out towards the high-mass star-forming region G34.26+0.15 with the HIFI instrument onboard the \textit{Herschel} Space Observatory, as well as with ground-based single-dish telescopes. Ten HDO lines and three H$_2^{18}$O lines covering a broad range of upper energy levels (22--204\,K) were detected. We used a non-LTE 1D analysis to determine the HDO/H$_2$O ratio as a function of radius in the envelope. Models with different water abundance distributions were considered in order to reproduce the observed line profiles. The HDO/H$_2$O ratio is found to be lower in the hot core ($\sim$3.5\,$\times$\,10$^{-4}$--7.5\,$\times$\,10$^{-4}$) than in the colder envelope ($\sim$1.0\,$\times$\,10$^{-3}$--2.2\,$\times$\,10$^{-3}$). This is the first time that a radial variation of the HDO/H$_2$O ratio has been found to occur in a high-mass source. The chemical evolution of this source was modeled as a function of its radius and the observations are relatively well reproduced. The comparison between the chemical model and the observations leads to an age of $\sim$10$^5$~years after the infrared dark cloud stage.
\end{abstract}


\section{Introduction}

Water, being necessary for the emergence of life, is one of the most important molecules found in space. As a dominant form of oxygen (the most abundant element 
in the Universe after hydrogen and helium), water controls the chemistry of many other species, whether in the gas phase or in the solid 
phase \citep[see for example the review by][]{vanDishoeck2013}. Water is a unique diagnostic of the warmer gas and the energetic processes taking place close to star-forming regions. 
Water is also a contributor to maintaining the low temperature of the gas by spectral line radiative cooling. 
Low temperatures are a requisite for cloud collapse and star formation. Water is mainly in its solid form (as ice on the surface of dust grains) in the cold 
regions of the interstellar medium as well as in asteroids and comets that likely delivered water to the Earth's oceans \citep[e.g.,][]{Hartogh2011,Alexander2012}. Therefore, 
constraining the distribution of water vapor and ice during the entire star and planet formation phase is mandatory to understand our own origins. 

Because of its high abundance in our own atmosphere, observations of interstellar water have been primarily carried out from space observatories including ISO, \textit{Spitzer}, 
ODIN, SWAS, and recently \textit{Herschel}. Indeed, water has been detected toward the cold prestellar core L1544 \citep{Caselli2012}, many low-mass 
protostars \citep[e.g.,][]{Coutens2012,Kristensen2010,Kristensen2012}, high-mass protostars 
\citep[e.g.,][]{vanderTak2013, Emprechtinger2013}, in the disk of a young star TW Hydrae \citep{Hogerheijde2011}, as well as in many comets (e.g., 103P/Hartley 2: \citealt{Hartogh2011}, 
C/2009 P1 (Garradd): \citealt{Bockelee2012}, 45P/Honda-Mrkos-Pajdu{\v s}{\'a}kov{\'a}: \citealt{Lis2013}) and in asteroids (24~Themis: \citealt{Campins2010}, Ceres: \citealt{Kuppers2014}). The water abundance shows a very large variation from one source category to another, as well as within each 
type of sources. The question then arises: how is water produced and why are its abundance variations so large? Although production in the gas phase followed by 
direct condensation onto dust grains is possible \citep{Bergin1999}, observations favor formation through chemical reactions on the surface of cold 
dust grains. Indeed, in comparison to gas-phase water abundance, the observed water ice abundance is too high to be entirely explained by direct accretion from 
the gas-phase \citep{Roberts2002}. Consequently surface reactions on cold dust grains to form water molecules have been investigated with modern 
surface science techniques \citep[e.g.][]{Watanabe2008}. Considering the large reservoir of oxygen and hydrogen atoms in molecular clouds, large amounts 
of water ice might be produced \citep{Dulieu2010} following the successive hydrogenation of oxygen on grain surfaces:
\begin{equation}
\rm O \xrightarrow{H} OH \xrightarrow{H} H_2O.
\end{equation}
\citet{Tielens1982} proposed that water ice might also be produced through the successive hydrogenation of molecular oxygen:
\begin{equation}
\rm O_2 \xrightarrow{H} HO_2 \xrightarrow{H} H_2O_2 \xrightarrow{H} H_2O + OH,
\end{equation}
demonstrated by \citet{Miyauchi2008}, \citet{Ioppolo2008} and \citet{Oba2009}, or by hydrogenation of ozone:
\begin{equation}
\rm O_3 \xrightarrow{H} O_2 + OH, \\
~~~~~~~~~~~~\rm OH \xrightarrow{H_2} H_2O + H,
\end{equation}
demonstrated by \citet{Mokrane2009}.

Deuterated water is likely to be formed through the same processes. Many rotational transitions have been detected 
from the ground, as well as with \textit{Herschel}/HIFI, for example in low-mass protostars \citep{Parise2005,Liu2011,Coutens2012,Coutens2013b,Persson2013,Persson2014}, high-mass star forming regions \citep[e.g.,][]{Jacq1990,Gensheimer1996}, and comets  \citep[e.g.,][]{Bockelee1998,Hartogh2011,Lis2013}. The HDO/H$_2$O ratio is an interesting diagnostic tool to help understand the origin of water in the 
interstellar medium, with a direct comparison with the D/H ratio observed in comets and in the Earth's oceans. It is also helpful to constrain the water formation conditions. In star-forming regions, observations of both high- and low-excitation water lines with a high spectral 
resolution are needed to disentangle the contributions from the hot cores (or hot corinos in the case of low-mass protostars) and the colder external envelope, 
that can be linked to the parental cloud, in which stars form. Near protostars, the grain temperature rises above $\sim$100\,K, leading to rapid water 
ice desorption that increases the gas-phase H$_2$O (and its deuterated counterparts) abundance in the inner parts of the envelope.
In order to interpret the observed spectra in terms of local physical conditions and relative abundances, radiative transfer modeling is necessary. This 
is illustrated with the modeling performed by \citet{Coutens2012,Coutens2013a} toward the low-mass protostar IRAS 16293-2422, where numerous HDO, H$_2^{18}$O, and D$_2$O transitions have been used simultaneously to constrain the abundances in the hot corino, in the cold envelope, and in a water-rich absorbing layer surrounding the envelope.

This paper reports full statistical equilibrium and radiative transfer calculations towards the ultra compact HII region G34.26+0.15 (hereafter G34) using both ground-based observations and 
\textit{Herschel}/HIFI observations of HDO and the less abundant H$_2^{18}$O water isotopologue. 
The paper is organized as follows. In Sections \ref{sect_source} and \ref{sect_obs}, we describe the source and the observations respectively.
In Section \ref{sect_rad_mod}, we present results obtained both with a simple local thermal equilibrium modeling (LTE) and with the 1D non-LTE modeling. In Section \ref{sect_chemistry}, we compare them with a chemical model. Finally, we present our conclusions in Section \ref{sect_conclu}.

\section{Source description}
\label{sect_source}

Located at a distance of $\sim$3.3 kpc \citep{Kuchar1994}, G34 has been widely studied in radio continuum \citep{Turner1974,Reid1985,Wood1989,Sewilo2011} and radio recombination lines 
\citep{Garay1985,Garay1986,Gaume1994,Sewilo2004,Sewilo2011}. Several components have been identified in radio continuum observations: two ultra compact 
HII regions called A and B, a more evolved HII region with a cometary shape (component C), and an extended (1$^{\prime}$) HII region (component D) in the 
south-east. Chemical surveys were carried out towards the A, B and C components using single-dish telescopes \citep{MacDonald1996,Hatchell1998} and interferometric observations \citep{Mookerjea2007}. 
Many complex species, characteristic of hot cores, have been detected. From molecular line observations, the emission peak does not coincide with the HII components \citep{Watt1999,DeBuizer2003}, but is shifted to the East of the component C by $\sim$1$^{\prime\prime}$ (\citealt{Mookerjea2007}: Figure 3). This difference may arise due to the external influence of the nearby HII regions, or may reveal separate regions of chemical enrichment. The hot core is likely externally heated by stellar photons rather than by shocks, as SiO was not detected at the position of the hot core \citep{Hatchell2001}. This source is also characterized by infall motions as suggested by observations of absorption components of NH$_3$, CN, HCN and HCO$^+$ \citep[][Hajigholi et al. in prep.]{Wyrowski2012,LiuT2013}.

The hot core of G34 has been the target of many \textit{Herschel}/HIFI observations for the past four years, including water line emission \citep{Flagey2013} and its deuterated couterparts. 
We present in Section \ref{sect_obs} the H$_2^{18}$O and HDO transitions observed from the ground as well as the \textit{Herschel}/HIFI observations. Note that the Half-Power Beam Width (HPBW) of those 
telescopes encompasses the components A and B and the molecular peak from component C for all observations.

\section{Observations}
\label{sect_obs}
 
 \begin{table*}
\centering
\caption{HDO and H$_2^{18}$O transitions observed towards the ultra-compact HII region G34$^{(1)}$. }
\label{table_obs_g34}
\begin{tabular}{ l c c c c c c c c c c c c}
\hline
\hline	
Species & Frequency & $J_{\rm Ka,Kc}$  & $E_{\rm up}/k$ & $A_{\rm ij}$ &  Telescope & HPBW & $F_{\rm eff}$ & $B_{\rm eff}$ & $d\varv$ & $rms^{(2)}$ & $\int T_{\rm mb} d\varv$  & $FWHM$ \\
 & (GHz) & & (K)  & (s$ ^{-1}$) &  & (\arcsec) & & & (km s$^{-1}$) &  (mK) & (K\,km s$^{-1}$) & (km s$^{-1}$) \\
\hline
HDO &  80.5783 & 1$_{1,0}$--1$_{1,1}$ & 47 & 1.32 $\times$ 10$^{-6}$ & IRAM-30m & 31.2 & 0.95 & 0.81 & 0.182 & 56 & 2.36 & 5.9  \\    
& 225.8967 & 3$_{1,2}$--2$_{2,1}$ & 168 & 1.32 $\times$ 10$^{-5}$ & IRAM-30m & 11.1 & 0.92 & 0.61 &  0.064 & 101&  10.45  & 6.7 \\    
& 241.5616 & 2$_{1,1}$--2$_{1,2}$ & 95 & 1.19 $\times$ 10$^{-5}$ & IRAM-30m & 10.4 & 0.90 & 0.56 & 0.061 & 84 & 12.27 & 6.6 \\    
& 464.9245 & 1$_{0,1}$--0$_{0,0}$ & 22 & 1.69 $\times$ 10$^{-4}$ & CSO & 16.5 & - & 0.35$^{(3)}$ & 0.078 & 304 & 7.48 & 5.2 \\    
 & 490.5966 & 2$_{0,2}$-1$_{1,1}$ & 66 & 5.25 $\times$ 10$^{-4}$ & HIFI 1a& 43.9 & 0.96 & 0.76 & 0.305 & 10 & 2.13 & 7.6\\    
& 509.2924 & 1$_{1,0}$--1$_{0,1}$ & 47 & 2.32 $\times$ 10$^{-3}$ & HIFI 1a& 42.3 & 0.96 & 0.76 & 0.294 & 44 & 1.76 &  9.0 \\    
 & 599.9267 & 2$_{1,1}$-2$_{0,2}$ & 95 & 3.45 $\times$ 10$^{-3}$ & HIFI 1b & 35.9 & 0.96 & 0.75 & 0.250 & 12 & 2.63 & 7.7 \\    
 & 848.9618 & 2$_{1,2}$-1$_{1,1}$ & 84 & 9.27 $\times$ 10$^{-4}$ & HIFI 3a & 25.4 & 0.96 & 0.75 &  0.176 & 10 & 3.92 & 10.3 \\  
& 893.6387 & 1$_{1,1}$--0$_{0,0}$ & 43 & 8.35 $\times$ 10$^{-3}$ & HIFI 3b & 24.1 & 0.96 & 0.74 & 0.167 & 63 & $-$2.38$^{(4)}$ & 5.9$^{(5)}$  \\  
 & 919.3109 & 2$_{0,2}$-1$_{0,1}$ & 66 & 1.56 $\times$ 10$^{-3}$ & HIFI 3b & 23.4 & 0.96 & 0.74 & 0.163 & 20 & 2.01 & 6.1  \\  
\hline
p--H$_2^{18}$O & 203.4075 & 3$_{1,3}$--2$_{2,0}$ & 204 & 4.81 $\times$ 10$^{-6}$ & IRAM-30m & 12.1 & 0.93 & 0.62 & 0.074 & 121 & 8.38$^{(6)}$ &  5.6$^{(6)}$\\
p--H$_2^{18}$O$^{(7)}$ & 1101.6983 & 1$_{1,1}$--0$_{0,0}$ & 53 & 1.79 $\times$ 10$^{-2}$ & HIFI 4b & 19.2 & 0.96 & 0.74 & 0.136 & 110 & $-$1.97$^{(8)}$ & 6.4 $^{(9)}$\\
o--H$_2^{18}$O$^{(7)}$ & 547.6764 & 1$_{1,0}$--1$_{0,1}$ & 60 & 3.29 $\times$ 10$^{-3}$ & HIFI 1a & 38.7 & 0.96 & 0.75 & 0.274 & 12 & 1.77$^{(10)}$ & 7.5$^{(11)}$\\
\hline
\end{tabular}
\begin{flushleft}

$^{(1)}$ The frequencies, upper energy levels ($E_{\rm up}$) and Einstein coefficients ($A_{\rm ij}$) come from the spectroscopic catalog JPL \citep{Pickett1998}. \\
$^{(2)}$ The $rms$ is calculated at the spectral resolution of the observations, which is indicated in the column $d\varv$.\\
$^{(3)}$ This value corresponds to the ratio between the main beam efficiency $B_{\rm eff}$ and the forward efficiency $F_{\rm eff}$.\\
$^{(4)}$ The integrated flux of the emission component is $\sim$ 0.87\,K.km s$^{-1}$, whereas it is $\sim$ $-$3.25\,K.km\,s$^{-1}$ for the absorbing component. \\
$^{(5)}$ The Full Width at Half Maximum ($FWHM$) of the fundamental line at 894 GHz is estimated to be 5.9 \kms~ for the emission component (v$_{LSR}$ = 58.0~\kms) and 3.9 \kms~for the absorption component (v$_{LSR}$ = 60.6 \kms). \\
$^{(6)}$ After subtraction of the CH$_3$OCH$_3$ line contaminating the para--H$_2^{18}$O line profile. \\
$^{(7)}$ Observations from \citet{Flagey2013}. \\
$^{(8)}$ The integrated flux of the emission component is $\sim$ 1.80\,K km s$^{-1}$, whereas it is $\sim$ $-$0.13\,K km\,s$^{-1}$ for the absorbing component. \\
$^{(9)}$ The Full Width at Half Maximum ($FWHM$) of the fundamental H$_2^{18}$O line at 1101 GHz is estimated to be 6.4 \kms~ for the emission component (v$_{LSR}$ = 57.5 \kms) and 3.5 \kms~for the absorption component (v$_{LSR}$ = 61.2 \kms). \\
$^{(10)}$ The integrated flux of the emission component is $\sim$ 1.39\,K km s$^{-1}$, whereas it is $\sim$ $-$3.36\,K km\,s$^{-1}$ for the absorbing component. \\
$^{(11)}$ The Full Width at Half Maximum ($FWHM$) of the fundamental H$_2^{18}$O line at 547 GHz is estimated to be 7.5 \kms~ for the emission component (v$_{LSR}$ = 57.3 \kms) and 3.3 \kms~for the absorption component (v$_{LSR}$ = 60.6 \kms).
\end{flushleft}
\end{table*}%

\subsection{Observations and data reduction}

 \begin{figure}
\begin{center}
\includegraphics[width=0.47\textwidth]{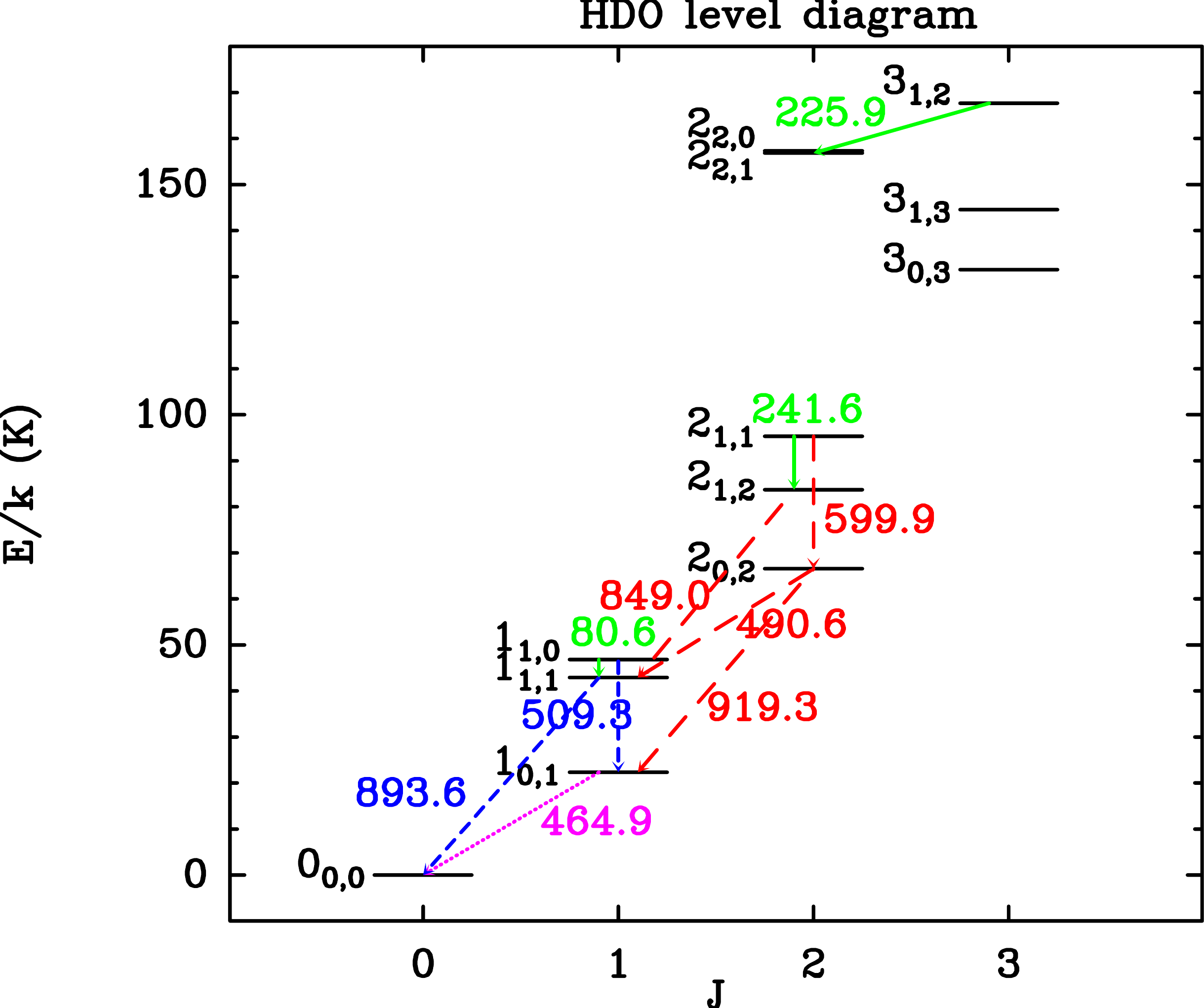}
\caption{Energy level diagram of the HDO lines. Green solid arrows: the IRAM-30m observations;
blue short dashed arrows: the PRISMAS/HIFI observations; red long dashed arrows: the Open Time HIFI observations; magenta dotted arrow: the CSO observation. The frequencies are given in GHz. (A color version of this figure is available in the online journal.)}
\label{nrjlevel}
\end{center}
\end{figure}

This source is part of the PRISMAS Key Program \citep[PRobing InterStellar Molecules with Absorption line Studies;][]{Gerin2010} which was followed by an Open Time Program led by C. Vastel. The targeted 
coordinates are $\alpha$(J2000) =  18$^h$53$^m$18.7$^s$, $\delta$(J2000) =
01$^\circ$14$^{\prime}$58$^{\prime\prime}$.  The
observations were performed in the pointed dual beam switch (DBS) mode
using the double sideband (DSB) HIFI instrument \citep{deGraauw2010,roelfsema2012} onboard the
{\it Herschel} Space Observatory \citep{pilbratt2010}. The DBS reference
positions were situated approximately 3$\arcmin$ east and west of the
source. The HIFI Wide Band Spectrometer (WBS) was used with optimization
of the continuum, providing a spectral resolution of 1.1\, MHz  over an
instantaneous bandwidth of 4~$\times$~1\,GHz. To disentangle the lines
of interest from the lines in the opposite sideband, possibly contaminating
our observations, we observed the same transition with 3 different Local Oscillator (LO)
settings. This method is necessary in such chemically rich regions in
order to ensure genuine detection of spectral lines.  The HDO data were
processed using the standard HIFI pipeline up to level 2 with the
ESA-supported package HIPE 8.0 \citep{ott2010} and were then exported as
FITS files into CLASS/GILDAS
format\footnote{\url{http://www.iram.fr/IRAMFR/GILDAS}} for subsequent data
reduction. The two linear polarizations were averaged to lower the noise in the
final spectrum. The baselines are well-fitted by straight lines
over the frequency range of the whole band and were subtracted from all observations. The single sideband
continuum temperature (that was obtained by dividing by 2 the DSB continuum derived from the linear fit obtained from line free regions in the spectrum, i.e. assuming a sideband gain ratio of unity) was added to the spectrum of the 1$_{1,1}$--0$_{0,0}$ fundamental line. 
To constrain the HDO/H$_2$O ratio, we also used two H$_2^{18}$O transitions observed in the framework of the PRISMAS program and previously published by \citet{Flagey2013}. We refer to this paper for the data reduction of these two lines.
A list of all the \textit{Herschel}/HIFI observations used in this paper is provided in Table A1. 

The ground state 1$_{0,1}$--0$_{0,0}$ HDO transition was observed at the Caltech Submillimeter Observatory (CSO) in September 2011 using the Fast Fourier Transform Spectrometer (FFTS) with 500 MHz bandwidth. The data were taken under good weather conditions, with 1.5 mm of precipitable water vapor. The beam switching mode has been used with a chop throw of 240$\arcsec$. The main beam efficiency was determined from total power observations of Mars. 
The system temperature was about 3500\,K during the run.
The single sideband continuum temperature ($\sim$\,2.2 K) was added to the final baseline-subtracted spectrum.

Three additional HDO transitions at 81 (1$_{1,0}$--1$_{1,1}$), 226 (3$_{1,2}$--2$_{2,1}$) and 242 GHz (2$_{1,1}$--2$_{1,2}$) as well as the ortho--H$_2^{18}$O transition at 203 GHz (3$_{1,3}$--2$_{2,0}$) were observed with the IRAM-30m telescope. The observations were carried out in December 2011 using the Fast Fourier Transform Spectrometer (FTS) at a 50 kHz resolution. 
The spectral resolution was 0.19, 0.07 and 0.06 $\kms$ for the 81, 226 and 242 GHz transitions, respectively. 
All the observations were performed using the Wobbler Switching mode. 
The 30m beam sizes at the observing frequencies are given in Table \ref{table_obs_g34}. During this run, weather conditions were good for winter, with 2 mm of precipitable water vapor. System temperatures were always less than 200\,K.

Figure \ref{nrjlevel} presents the energy level diagram of the HDO transitions used for the modeling. Table \ref{table_obs_g34} summarizes the observations.

\subsection{Description of the observations}

Most of the observed HDO lines show a Gaussian-like profile (see for example Fig. \ref{ratran_hdo_100K}).
Only the HDO 1$_{1,1}$--0$_{0,0}$ fundamental transition observed with \textit{Herschel}/HIFI shows an inverse P-Cygni profile, i.e a profile showing a red-shifted absorption component and a blue-shifted emission component.
A similar profile has already been observed for this transition in low-mass protostars \citep{Coutens2012,Coutens2013b}.
The Gaussian FWHM (Full Width at Half-Maximum) was derived for each line with the CASSIS\footnote{\url{http://cassis.irap.omp.eu}} software (see Table \ref{table_obs_g34}).
Using the available spectroscopic databases CDMS (Cologne Database Molecular Spectroscopy; \citealt{Muller2011,Muller2005}) and JPL (Jet Propulsion Laboratory; \citealt{Pickett1998}), we also checked that the different lines are not contaminated by other species.
The HDO 2$_{1,1}$--2$_{1,2}$ transition at 242 GHz could be slightly blended with the CH$_3$COCH$_3$ 13$_{10,4}$--12$_{9,3}$ line. However a simple LTE (Local Thermal Equilibrium) modeling of CH$_3$COCH$_3$ lines observed in the spectra, shows that the contribution of CH$_3$COCH$_3$ is negligible. With a column density of 4\,$\times$\,10$^{16}$ cm$^{-2}$, an excitation temperature of 100\,K, a FWHM of 6 $\kms$ and a source size of 1.7$\arcsec$, the predicted intensity of the CH$_3$COCH$_3$ line at 241.6 GHz is 0.06\,K, to be compared with the observed line intensity 1.75\,K. 
The HDO 2$_{1,2}$--1$_{1,1}$ line at 849 GHz is very probably blended with three $^{13}$CH$_3$OH lines (18$_{3,15}$--17$_{3,14}$ A-, 18$_{4,14}$--17$_{4,13}$ A+, and 18$_{4,15}$--17$_{4,14}$ A-) lying in the red-shifted portion of the line profile. This could explain why this line is broader than the others (see Table \ref{table_obs_g34}). As the $^{13}$CH$_3$OH contribution could be non-negligible, we do not use this HDO line to constrain the abundances. We present however the modeling of this line for completeness.
The other HDO lines do not show any potential blending. The portion of the 600 GHz line observed at $\varv$ $>$ 72 $\kms$ is produced by the CH$_3$OH  7$_{3,5}$--6$_{2,4}$ $\varv$=0 A+ line from the image band ($\nu$ = 590.3 GHz).

 \begin{figure}
\begin{center}
\includegraphics[width=0.5\textwidth]{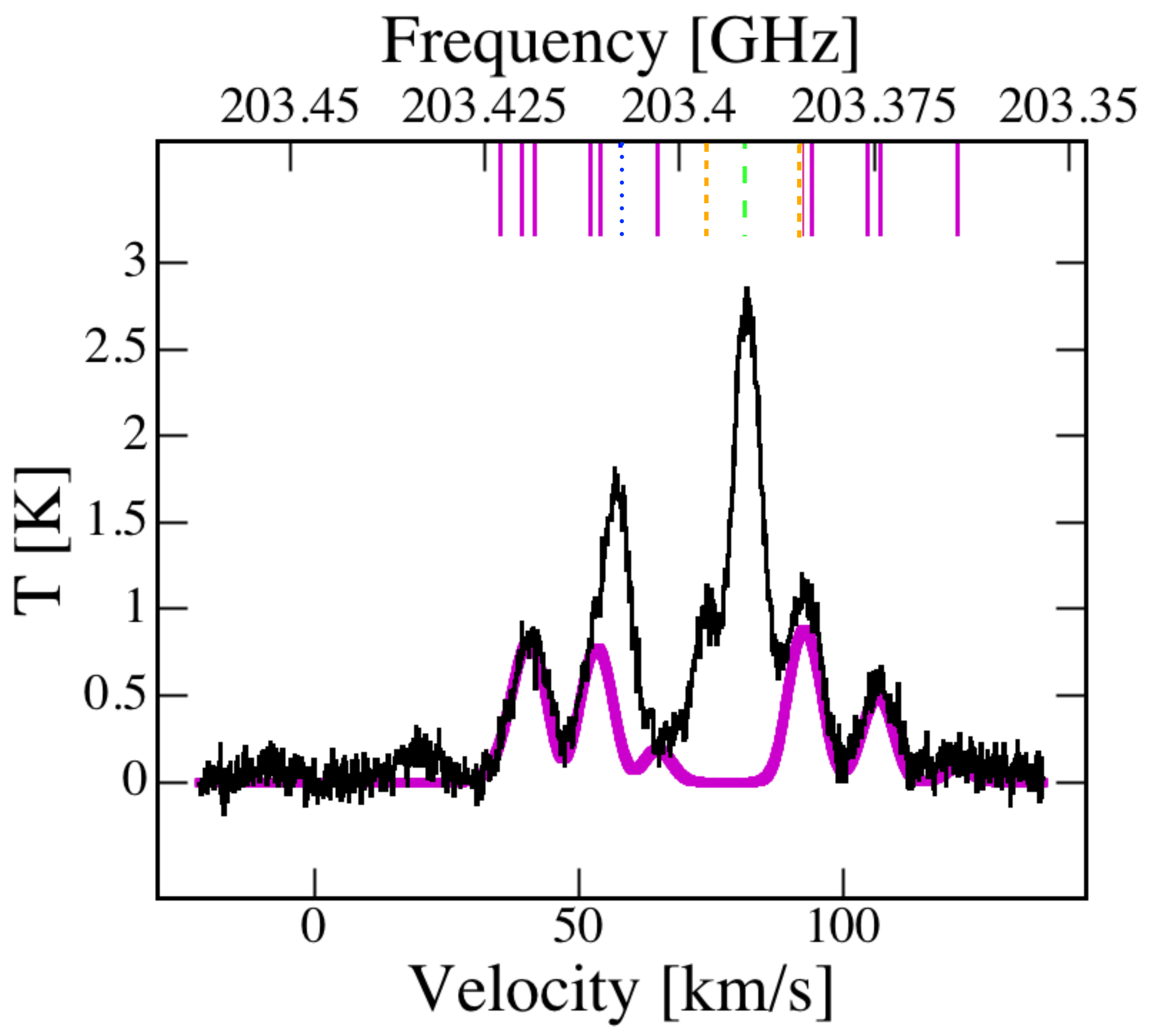}
\caption{IRAM-30m observations (in black) of the para--H$_2^{18}$O 3$_{1,3}$--2$_{2,0}$ transition at 203.4 GHz. The frequency of the H$_2^{18}$O line is indicated by a blue dotted line ($\varv$ = 58 $\kms$). The other lines observed in this spectra are the SO$_2$ $\varv$=0 12$_{0,12}$--11$_{1,11}$ line at 81.5 $\kms$ (green long dashed line), the C$_2$H$_5$CN $\varv$=0 23$_{2,22}$--22$_{2,21}$ line at 74.1 $\kms$ (yellow short dashed line) and several CH$_3$OCH$_3$ lines (magenta solid lines).
A LTE modeling of the CH$_3$OCH$_3$ lines (in magenta) was carried out to estimate the contamination of the para--H$_2^{18}$O line by the CH$_3$OCH$_3$ 3$_{3,1,1}$--2$_{2,1,1}$ transition. (A color version of this figure is available in the online journal.)
}
\label{ch3och3_blend}
\end{center}
\end{figure}

The para--H$_2^{18}$O 3$_{1,3}$--2$_{2,0}$ line is blended with the CH$_3$OCH$_3$ 3$_{3,1,1}$--2$_{2,1,1}$ and 3$_{3,0,3}$--2$_{2,1,3}$ transitions at 203.4101 and 203.4114\,GHz ($E_{\rm up}$ = 18\,K).
We can reproduce, with a LTE modeling approach, the CH$_3$OCH$_3$ lines observed nearby in the spectra (see Figure \ref{ch3och3_blend}) as well as in the other bands. The CH$_3$OCH$_3$ lines are well-fitted with a column density of 7\,$\times$\,10$^{17}$ cm$^{-2}$, an excitation temperature of 100\,K, a FWHM of 6 $\kms$ and a source size of 1.7$\arcsec$. 
The predicted line profiles of the CH$_3$OCH$_3$ transitions blended with H$_2^{18}$O are then subtracted from the observed line profile to extract the proper H$_2^{18}$O spectrum. Due to the high number of CH$_3$OCH$_3$ lines considered in the analysis and the presence of CH$_3$OCH$_3$ lines with similar upper energy levels (18\,K) around the H$_2^{18}$O feature (see Figure \ref{ch3och3_blend}), the uncertainty produced by this subtraction is negligible with respect to the calibration uncertainty ($\leq$20\%).

\section{Radiative transfer modeling}
\label{sect_rad_mod}

\subsection{Rotational diagram analysis}
\label{sect_lte}

\begin{figure}
\begin{center}
\includegraphics[width=0.45\textwidth]{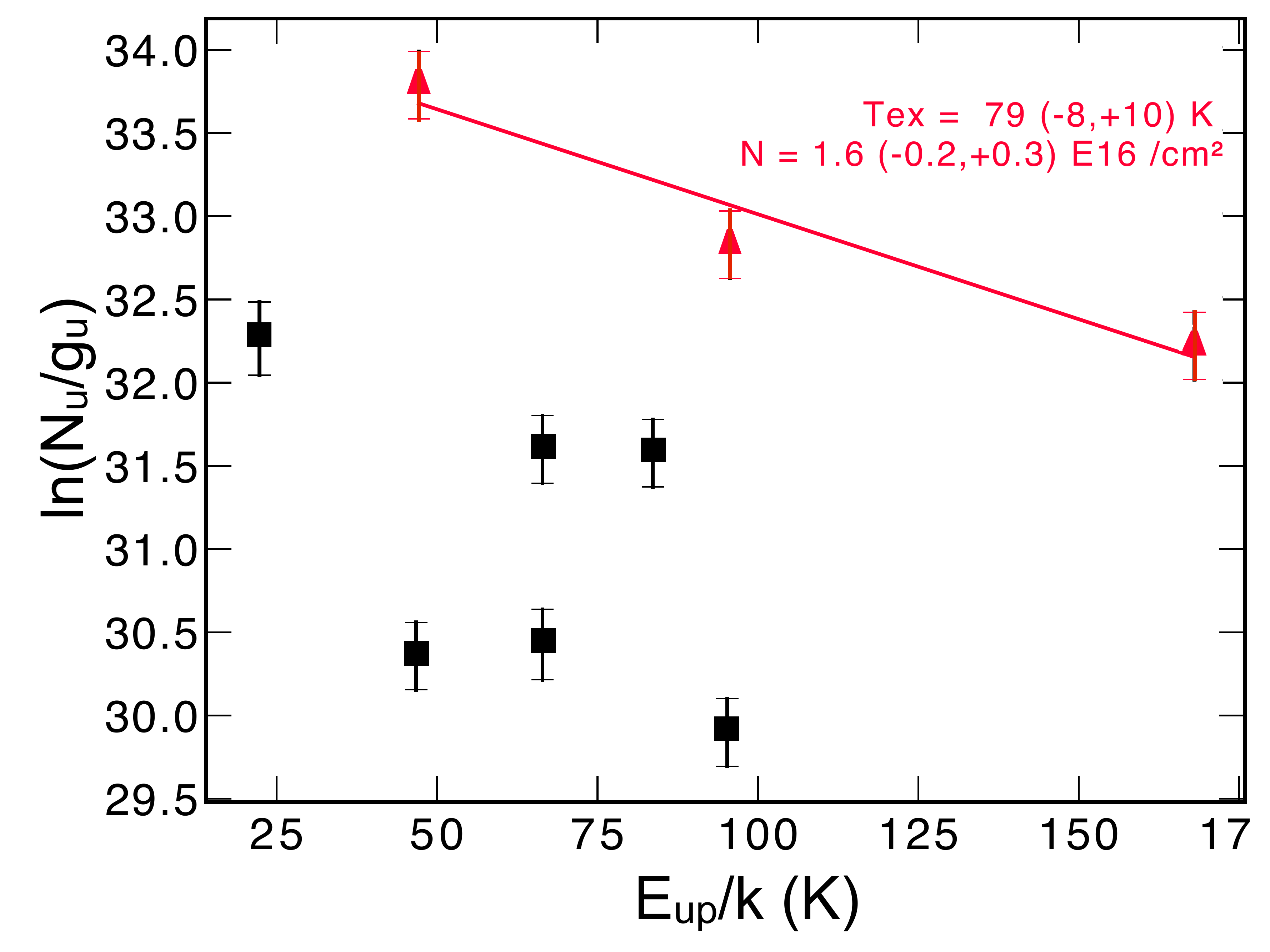}
\caption{Rotational diagram of the HDO lines. The fundamental line at 894 GHz is excluded from the figure. A source size of 4$\arcsec$ is assumed. The error bars correspond to uncertainties of 20\%. A linear fit was made using only the lines with low critical densities (81, 226, and 242 GHz, indicated by red triangles) to estimate the HDO column density ($\sim$\,1.6\,$\times$\,10$^{16}$ cm$^{-2}$) and the excitation temperature ($\sim$\,79\,K) in the hot core (see text). (A color version of this figure is available in the online journal.)}
\label{RD_HDO}
\end{center}
\end{figure}

A simple LTE modeling was first employed to estimate the HDO/H$_2$O ratio in the hot core.
We plot in Figure \ref{RD_HDO} the rotation diagram \citep{Goldsmith1999} of the HDO lines shown in Table \ref{table_obs_g34}. We exclude the fundamental transition at 894 GHz, which shows absorption and probably probes colder regions outside of the hot core.
We take into account beam dilution and consider different source sizes between 1$\arcsec$ and 5$\arcsec$, as the exact size of the hot core is unknown. Indeed, the structure determined by \citet{vanderTak2013} predicts a size of 4.5$\arcsec$ for T $>$ 100\,K, whereas the interferometric observations of two HDO lines by \citet{Liu2013} favor a smaller source size which, however, is not well constrained.
No linear curve is in reasonable agreement with the complete dataset. Plausible explanations are that the lines are optically thick or that they do not all have the same excitation temperature.
We estimate the critical densities of these species using the HDO collisional coefficients with ortho-- and para--H$_2$ of \citet{Faure2011}. At $\sim$100\,K, the critical densities are about 5\,$\times$\,10$^{6}$ -- 5\,$\times$\,10$^{7}$ cm$^{-3}$ for all lines, except for the lines at 80, 226, and 242 GHz that have critical densities between 3\,$\times$\,10$^{4}$ and 3\,$\times$\,10$^{5}$ cm$^{-3}$. These latter lines are at low frequencies, so that their radiative decay is slower. They are consequently probably in LTE, as the density in the hot core is expected to be $\gtrsim$ 10$^6$ cm$^{-3}$ \citep{vanderTak2013}. In addition, these three lines are also those with the expected lowest opacities. 
We consequently fit a straight line to these three points only. The column density and the excitation temperature of HDO were then estimated for different values of the source size (1$\arcsec$--5$\arcsec$). To derive the HDO/H$_2$O ratio, we used the H$_2^{18}$O 3$_{1,3}$--2$_{2,0}$ line observed at 203 GHz with IRAM. Indeed this line is quite excited and its critical density is relatively low, about 10$^{5}$ cm$^{-3}$. Using the same excitation temperature as HDO, we calculated the column density of H$_2^{18}$O in the hot core and derived an estimate of the HDO/H$_2$O ratio between 5.2\,$\times$\,10$^{-4}$ and 5.7\,$\times$\,10$^{-4}$. The H$_2^{16}$O/H$_2^{18}$O ratio is assumed to be 400 following the relation determined by \citet{Wilson1999} between the $^{16}$O/$^{18}$O isotopic ratio and the distance of the source from the galactic center. The derived HDO/H$_2$O ratio is consistent with the previous estimates by \citet[][4\,$\times$\,10$^{-4}$]{Jacq1990} and \citet[][3.0\,$\times$\,10$^{-4}$]{Liu2013}, who assumed an H$_2^{16}$O/H$_2^{18}$O ratio equal to 500. It is also slightly greater than the estimate by \citet[][1.1\,$\times$\,10$^{-4}$]{Gensheimer1996}.

\subsection{Non-LTE spherical modeling }
\label{sect_modeling}

\subsubsection{Model description}

Only three HDO lines (among 10) were used to estimate the HDO/H$_2$O ratio in the hot core with the rotational diagram analysis. We consequently decided to employ non-LTE spherical modeling (that also considers the line opacities) to include the information provided by all the HDO lines (except the 848 GHz line that is probably blended with $^{13}$CH$_3$OH) and to determine the water deuterium fractionation in both the hot core and the colder part of the envelope. 
We used the RATRAN code \citep{ratran} that assumes spherical symmetry and takes into account continuum emission and absorption by dust.
To derive the HDO and H$_2^{18}$O abundances, we used the temperature and H$_2$ density profiles derived by \citet[Section 4.1]{vanderTak2013}.
This structure was determined taking into account JCMT/SCUBA and PACS data. 
The radial velocity profile ($\varv_r$) and the turbulence width (Doppler b-parameter, $db$) have also to be provided in RATRAN.  
We describe in Appendix B2 the method employed to constrain them and show the final $\varv_r$ and $db$ profiles used in the analysis. We find that inward motions ($\varv_r$~$\sim$~-3 km\,s$^{-1}$) are present in the cold envelope, while outward motions ($\varv_r$~$\sim$~4 km\,s$^{-1}$) take place in the inner regions. The same type of velocity profile was found in SgrB2(M) by \citet{Rolffs2010}. The Doppler b-parameter appears lower in the inner regions ($db$~$\sim$~2.0 km\,s$^{-1}$) than in the outer regions ($db$~$\sim$~2.5 km\,s$^{-1}$), similarly to what was found by \citet{Caselli1995} and \citet{Herpin2012} in other high mass sources.
To reproduce the continuum levels seen in the observations as best as possible, we used the dust opacities from \citet{Ossenkopf1994}, with thick ice mantles and a gas density of 10$^6$ cm$^{-3}$. The dust opacities used by \citet[][thin ices mantles with gas density of 10$^6$ cm$^{-3}$]{vanderTak2013} to derive the structure would not however differ too much, as the predicted continuum is consistent with the observations to within 10--20\% uncertainties.
The most recent HDO and H$_2^{18}$O collisional coefficients calculated with ortho-- and para--H$_2$ by \citet{Faure2011} and \citet{Daniel2011}, respectively, were used. The ortho/para ratio of H$_2$ is assumed to be at LTE in each cell of the envelope. It consequently varies from $\sim$10$^{-2}$ in the coldest regions up to the equilibrium value of 3 in the warm regions.

\subsubsection{Modeling of the HDO lines with an abundance jump}
\label{sect_jump}

\begin{figure*}
\begin{center}
\includegraphics[width=1.0\textwidth]{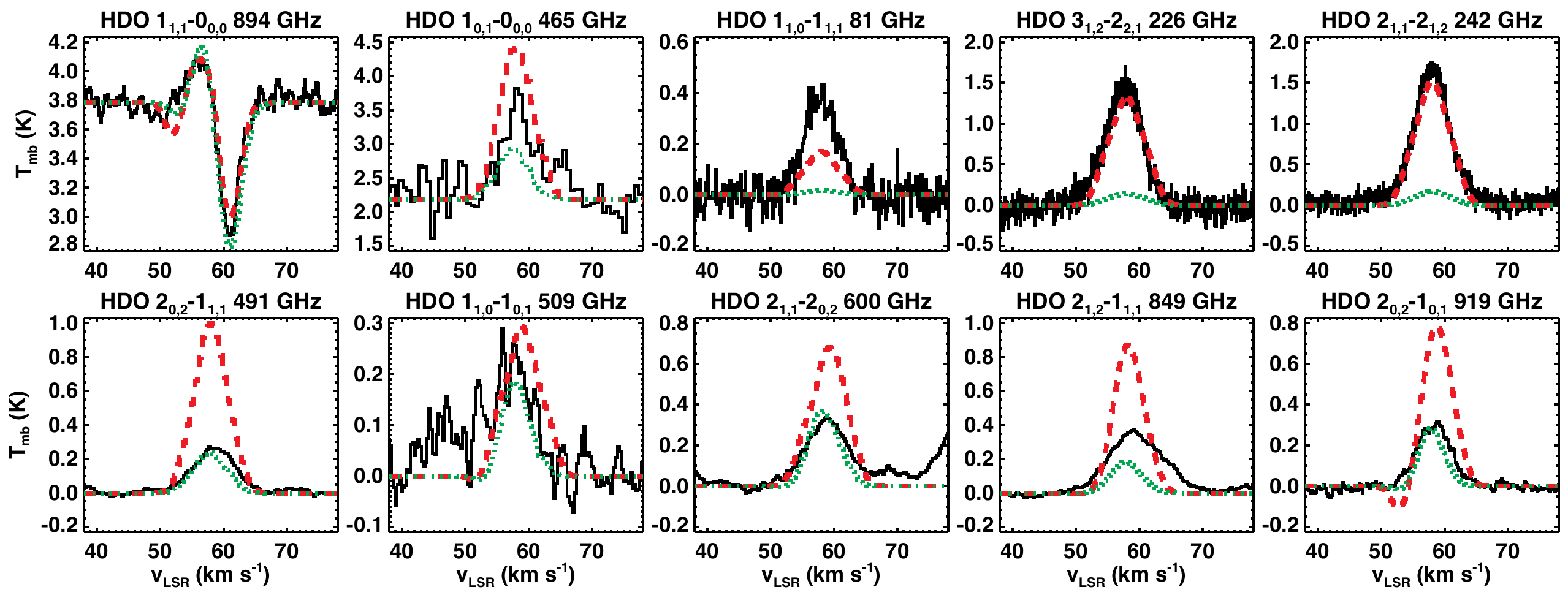}
\caption{{\it Black solid line}: HDO lines observed with HIFI, IRAM, and CSO.
{\it Red dashed line}: Modeling for a jump temperature $T_{\rm j}$ = 100\,K, an inner abundance $X_{\rm in}$ = 3\,$\times$\,10$^{-8}$ and an outer abundance $X_{\rm out}$ = 8\,$\times$\,10$^{-11}$.
{\it Green dotted line}: Modeling for a jump temperature $T_{\rm j}$ = 100\,K, an inner abundance $X_{\rm in}$ = 3\,$\times$\,10$^{-9}$ and an outer abundance $X_{\rm out}$ = 8\,$\times$\,10$^{-11}$. (A color version of this figure is available in the online journal.)}
\label{ratran_hdo_100K}
\end{center}
\end{figure*}

Most of the studies of water and deuterated water in star-forming regions \citep[e.g.,][]{Ceccarelli2000,Parise2005,vandertak2006,Coutens2012,Herpin2012} assume an abundance jump at $T_{\rm j}$ = 100\,K, corresponding to the temperature at which the water molecules are supposed to be released in the gas phase by thermal desorption. In a first step, we consequently assumed such an abundance jump for the modeling of the HDO lines. According to the physical structure used here \citep{vanderTak2013}, the source size corresponding to $T$ $>$ 100\,K is $\sim$4.5$\arcsec$ (diameter).
We ran a grid of models with various inner ($T > 100$\,K) and outer ($T < 100$\,K) abundances and realized that, regardless of the velocity profiles, the intensities of the different lines cannot be reproduced simultaneously (see Figure \ref{ratran_hdo_100K}). 
Indeed, when the excited transitions observed at 225 and 241 GHz with IRAM are reproduced, the fluxes of the CSO and HIFI lines are overproduced, in particular at 491, 600, 849, and 919 GHz (red dashed model in Figure \ref{ratran_hdo_100K}).
On the contrary, if these latter lines are reproduced, the flux is underpredicted for the IRAM lines (green dotted model in Figure \ref{ratran_hdo_100K}). 
Although the choice of the velocity profiles can affect the line profiles, it is not possible to appreciably modify the intensities and decrease this disagreement. 

\begin{figure*}
\begin{center}
\includegraphics[width=1.0\textwidth]{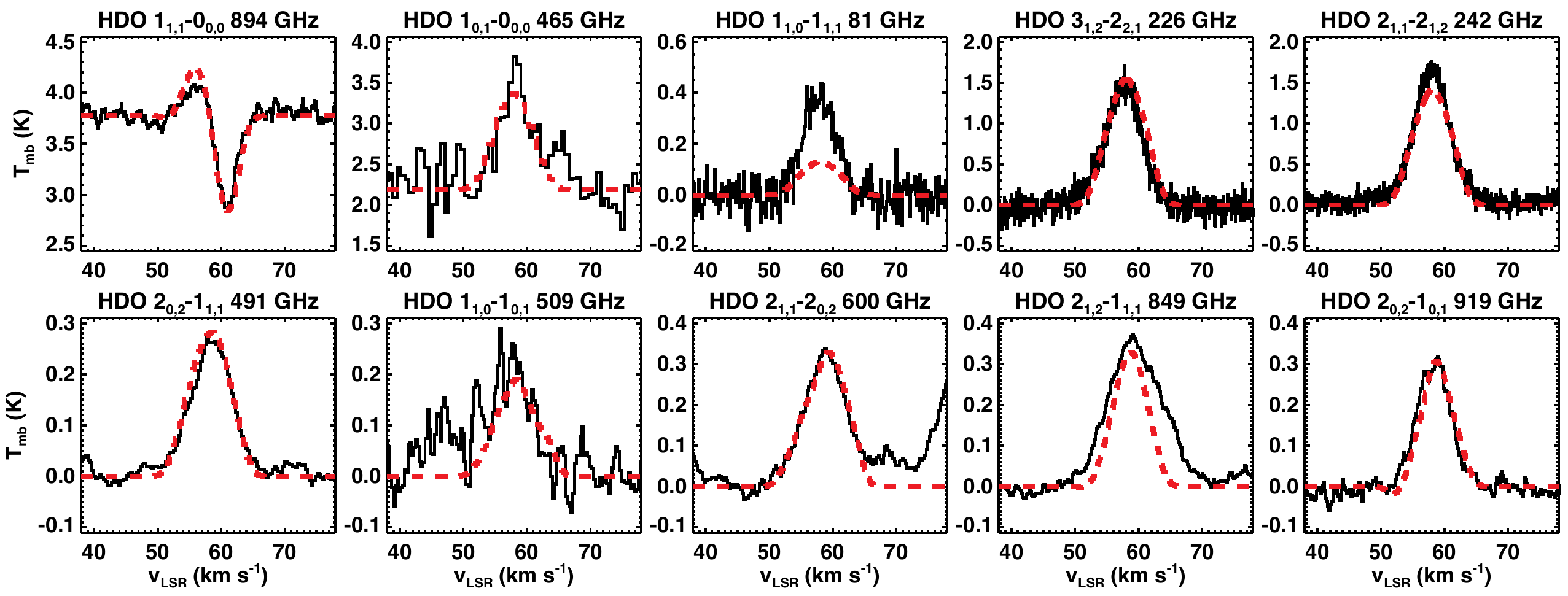}
\caption{{\it Black solid line}: HDO lines observed with HIFI, IRAM, and CSO.
{\it Red dashed line}: Modeling for a jump temperature $T_{\rm j}$ = 200 K, an inner abundance $X_{\rm in}$ = 2\,$\times$\,10$^{-7}$ and an outer abundance $X_{\rm out}$ = 8\,$\times$\,10$^{-11}$. (A color version of this figure is available in the online journal.)}
\label{ratran_hdo_200K}
\end{center}
\end{figure*}

To obtain an agreement for all the transitions, an increase of the jump temperature is necessary.
Indeed, Figure B4 shows that, with a jump temperature of 120\,K, the model that reproduces the fluxes of the most excited HDO lines (226 and 242 GHz) is in better agreement with the fluxes of the lines at 491, 600, 849, and 919 GHz than the model with a jump at $T_{\rm j}$ = 100\,K. The fluxes of these four lines are, however, still overproduced.
Consequently, we ran grid of models for higher jump temperatures (150, 180, 200, and 220\,K) and compared the influence of the jump temperature on the line intensities. 
The best-fit predictions obtained for $T_{\rm j}$~=~150\,K, 180\,K, 200\,K, and 220\,K are shown in Figures B5, B6, \ref{ratran_hdo_200K}, and B7, respectively. These four models reproduce relatively well the observations. The 491, 600, 849, and 919 GHz lines are quite sensitive to the jump temperature. Their intensities decrease with the increase of $T_{\rm j}$. In particular, for $T_{\rm j}$~=~150\,K and 180\,K, their intensities are slightly overpredicted, while, for $T_{\rm j}$ $\geq$ 220\,K, they start to be under-predicted. In view of these results, the best-fit model is obtained for $T_{\rm j}$ $\sim$\,200\,K. 
Table \ref{table_results} summarizes the HDO best-fit abundances found for each jump temperature and the corresponding size of the jump abundance. 
The best-fit was determined with a $\chi^2$ minimization of the line profiles similar to what was done in \citet{Coutens2012}, assuming a calibration uncertainty of 20\% for each line. As the HDO line at 849\,GHz is probably blended with $^{13}$CH$_3$OH, we did not include it in the calculation. The reduced $\chi^2$ obtained for the model with an abundance jump at 200\,K is 1.3. The HDO inner abundance is strongly constrained by the high number of emission lines used in the analysis. If we just consider the grid with $T_{\rm j}$~=~200\,K, its value is between 1.7\,$\times$\,10$^{-7}$ and 2.1\,$\times$\,10$^{-7}$. Consequently, the main uncertainty on the HDO inner abundance comes from the value assumed for the jump temperature ($\sim$1$\times$\,10$^{-7}$--3\,$\times$\,10$^{-7}$ for $T_{\rm j}$$\sim$150--220\,K). The outer abundance is mainly constrained by the absorbing component at 894 GHz and its uncertainty is found to be between 6\,$\times$\,10$^{-11}$ and 9\,$\times$\,10$^{-11}$.
The HDO 1$_{1,0}$--1$_{1,1}$ line at 81 GHz is not reproduced by any of the models within the 20\% calibration uncertainty and could maybe suffer of calibration problems at this low frequency with the IRAM-30m telescope. Models with a two-jump abundance profile such as in \citet{Comito2010} were also attempted but do not improve the fit (see Appendix B3).

It clearly appears that, to reproduce the HDO line profiles, an increase of the jump temperature in the model is necessary.
We cannot conclude, however, that the sublimation temperature for water ice is significantly higher than 100\,K. Although some experiments actually favor an evaporation temperature of 110--120\,K \citep{Fraser2001}, it is not sufficient to perfectly reproduce all the HDO transitions. 
The main reason for the modification of the jump temperature would be related to the size of the hot core, rather than to the water sublimation temperature itself.
In this case, the size of the hot core in which the abundance of water increases after the evaporation of the icy mantles should be smaller ($\sim$2$\arcsec$ instead of 4.5$\arcsec$), in order to lead to a better agreement between the model and the observations. This is also in agreement with the interferometric observations of the HDO lines at 225 and 241 GHz by \citet{Liu2013} that are not spatially resolved with a beam size of 3.7$\arcsec$\,$\times$\,2.5$\arcsec$. Two explanations can be provided to explain the smaller size of the hot core.
One would be that the physical structure derived by \citet{vanderTak2013} is unreliable at small scales.
Indeed the structure determined here is only based on large-scale maps and the density profile is assumed to follow a power-law. The density and temperature profiles could therefore be uncertain at small scales ($\theta$ $\lesssim$ 5$\arcsec$). In this case, the temperature actually would reach 100\,K at a radius which is smaller than what the physical structure predicts \citep{vanderTak2013}.
The second possible explanation is provided by the chemical models coupled with a dynamical approach, where the dynamical timescales can be in competition with the chemical and adsorption/desorption timescales. Indeed, as it can be seen for example in \citet{Aikawa2012} and Wakelam et al. (submitted), the abundance increases gradually for a certain temperature range before a constant inner abundance is reached. The temperature where the inner abundance is constant is higher than 100\,K but its exact value is dependent on the model parameters. It seems therefore possible that the constant inner abundance can be reached only at $\sim$200\,K.
Some tests assuming a gradual abundance increase were attempted in Section \ref{sect_mod_increase} and this explanation seems to hold here.

\begin{figure*}
\begin{center}
\includegraphics[width=0.62\textwidth]{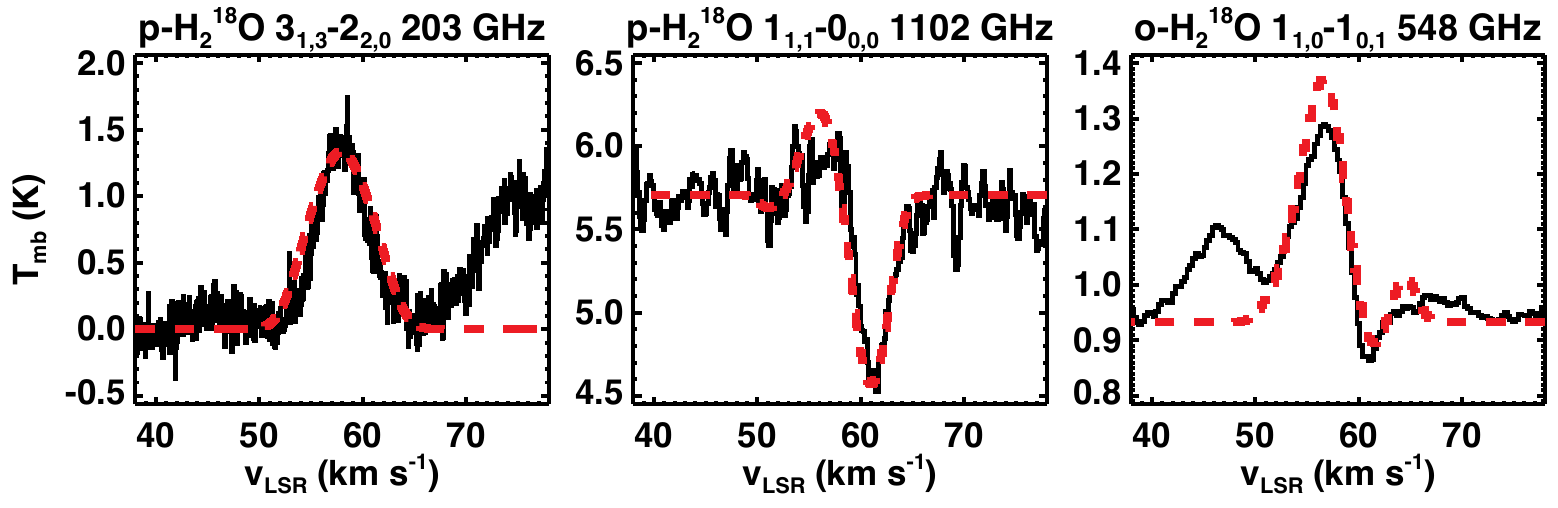}
\caption{{\it Black solid line}: H$_2^{18}$O lines observed with HIFI and IRAM.
{\it Red dashed line}: Modeling for a jump temperature $T_{\rm j}$ = 200 K, an inner abundance $X_{\rm in}$~=~9\,$\times$\,10$^{-7}$ and an outer abundance $X_{\rm out}$ = 1.3\,$\times$\,10$^{-10}$. (A color version of this figure is available in the online journal.)}
\label{ratran_h218o_200K}
\end{center}
\end{figure*}
 
\begin{table*}
\caption{HDO and H$_2$O abundances obtained for different jump temperatures $T_{\rm j}$}
\begin{center}
\begin{tabular}{l |c |c c |c c| c c| c c}
\hline
\hline
$T_{\rm j}$ (K) $^{(a)}$ & $\theta$ ($\arcsec$) $^{(a)}$ & $X_{\rm in}$(HDO) & $X_{\rm out}$(HDO) & $X_{\rm in}$(H$_2^{18}$O) & $X_{\rm out}$(H$_2^{18}$O) & $X_{\rm in}$(H$_2$O)$^{(b)}$ & $X_{\rm out}$(H$_2$O)$^{(b)}$ & (HDO/H$_2$O)$_{\rm in}$$^{(b)}$ & (HDO/H$_2$O)$_{\rm out}$$^{(b)}$ \\
\hline
100$^{(c)}$ & 4.5 & -- & -- & -- & -- & -- & -- & -- & -- \\
120$^{(c)}$ & 3.5 & -- & -- & -- & -- & -- & -- & -- & -- \\
150 & 2.5 & 1 $\times$ 10$^{-7}$ & 8 $\times$ 10$^{-11}$ &  4 $\times$ 10$^{-7}$ & 1.3 $\times$ 10$^{-10}$ & 1.6 $\times$ 10$^{-4}$ & 5.2 $\times$ 10$^{-8}$ & 6 $\times$ 10$^{-4}$ & 1.5 $\times$ 10$^{-3}$  \\
180 & 1.9 & 1.5 $\times$ 10$^{-7}$ & 8 $\times$ 10$^{-11}$ & 7 $\times$ 10$^{-7}$ & 1.3 $\times$ 10$^{-10}$ & 2.8 $\times$ 10$^{-4}$ & 5.2 $\times$ 10$^{-8}$ & 5 $\times$ 10$^{-4}$ & 1.5 $\times$ 10$^{-3}$ \\
\bf 200$^{(d)}$ & \bf 1.7 & \bf 2 $\times$ 10$^{-7}$ & \bf 8 $\times$ 10$^{-11}$ & \bf 9 $\times$ 10$^{-7}$ & \bf 1.3 $\times$ 10$^{-10}$ & \bf 3.6 $\times$ 10$^{-4}$ & \bf 5.2 $\times$ 10$^{-8}$ & \bf 6 $\times$ 10$^{-4}$ & \bf 1.5 $\times$ 10$^{-3}$ \\
220 & 1.5 & 3 $\times$ 10$^{-7}$ & 8 $\times$ 10$^{-11}$ & 1.2 $\times$ 10$^{-6}$ & 1.3 $\times$ 10$^{-10}$ & 4.8 $\times$ 10$^{-4}$ & 5.2 $\times$ 10$^{-8}$ & 6 $\times$ 10$^{-4}$ & 1.5 $\times$ 10$^{-3}$  \\
\hline
\end{tabular}
\end{center}
{\footnotesize 
Notes: $^{(a)}$ Size of the region where the temperature is higher than $T_{\rm j}$ (diameter). It is derived from the structure determined by \citet{vanderTak2013}. $^{(b)}$ Assuming H$_2^{16}$O/H$_2^{18}$O\,=\,400. $^{(c)}$ Fit is not good enough to determine the HDO abundances. $^{(d)}$ Best-fit.}
\label{table_results}
\end{table*}%

\subsubsection{Modeling of the H$_2^{18}$O lines with an abundance jump and estimate of the HDO/H$_2$O ratios}
\label{sect_h218o}

A similar model with an abundance jump was carried out with the H$_2^{18}$O lines detected in this source to determine the HDO/H$_2$O ratio throughout the envelope. The H$_2^{16}$O transitions detected with the HIFI instrument by \citet{Flagey2013} are not well suited to measure abundances, because of their large opacities.
With its excitation level, the para--H$_2^{18}$O 3$_{1,3}$--2$_{2,0}$ transition observed at 203 GHz with the IRAM-30m telescope is suitable to probe the hot core and derive the HDO/H$_2$O ratio in the warm inner region. The H$_2^{18}$O fundamental lines previously detected with \textit{Herschel}/HIFI by \citet{Flagey2013} are also used to constrain the HDO/H$_2$O ratio in the envelope, as these lines combine both emission and absorption. Note that we only use here the ortho 1$_{1,0}$--1$_{0,1}$ transition at 548 GHz and the para 1$_{1,1}$--0$_{0,0}$ transition at 1102 GHz. The fundamental ortho 2$_{1,2}$--1$_{0,1}$ transition, which is observed at 1656 GHz in absorption, was not taken into account because of pointing problems affecting the observations. The source being fairly peaked on the continuum, an offset could lead to a significant loss of flux. 

All the physical parameters are kept similar to those of the study of HDO.
Figures B8, B9, \ref{ratran_h218o_200K}, and B10 show the best-fit models obtained for these three lines for the jump temperatures previously assumed for deuterated water, $T_{\rm j}$ = 150, 180, 200, and 220\,K respectively. 
We assumed an ortho-to-para ratio of water equal to 3, corresponding to the thermal equilibrium value at high-temperature ($>$ 50\,K). This value is also consistent with the ratio determined in most of the foreground clouds on the line of sight towards bright continuum sources \citep{Lis2010,Flagey2013}.
The best-fit inner and outer H$_2^{18}$O abundances are summarized in Table \ref{table_results}. The reduced $\chi^2$ is about 1.5 for the case $T_{\rm j}$\,=\,200\,K.
 Assuming an observational uncertainty of 20\% for the excited para--H$_2^{18}$O line at 203 GHz, the inner abundance cannot be higher than 1.2\,$\times$\,10$^{-6}$ or lower than 7\,$\times$\,10$^{-7}$ for the model with $T_{\rm j}$\,=\,200\,K.
The outer abundance is estimated to be between 1.0\,$\times$\,10$^{-10}$ and 1.5\,$\times$\,10$^{-10}$, based on an observational uncertainty of 20\% for the absorbing component of the para--H$_2^{18}$O transition at 1102 GHz.
The H$_2^{16}$O abundances in Table \ref{table_results} are estimated using an H$_2^{16}$O/H$_2^{18}$O ratio of 400 \citep{Wilson1999}.

The best-fit HDO/H$_2$O ratios are then equal to $\sim$(5--6)\,$\times$\,10$^{-4}$ in the hot core and $\sim$1.5\,$\times$\,10$^{-3}$ in the outer envelope. Even when considering the HDO and H$_2^{18}$O results with a 20\% calibration uncertainty, the outer HDO/H$_2$O ratio (1.0\,$\times$\,10$^{-3}$--2.2\,$\times$\,10$^{-3}$) is still higher than the inner HDO/H$_2$O ratio (3.5\,$\times$\,10$^{-4}$--7.5\,$\times$\,10$^{-4}$ for $T_{\rm j}$ = 200\,K).
We ran models with a constant ortho/para H$_2$ ratio equal to 3 to check that the ortho/para H$_2$ ratio assumed in the model does not affect the results. The HDO and H$_2^{18}$O line profiles are exactly the same as with an LTE ortho/para ratio, confirming the variation of the HDO/H$_2$O ratio from the cold to the warm regions.

\subsubsection{Gradual decrease of the outer abundance from the cold to the warm regions}
\label{sect_slope}

\begin{figure}
\begin{center}
\includegraphics[width=0.5\textwidth]{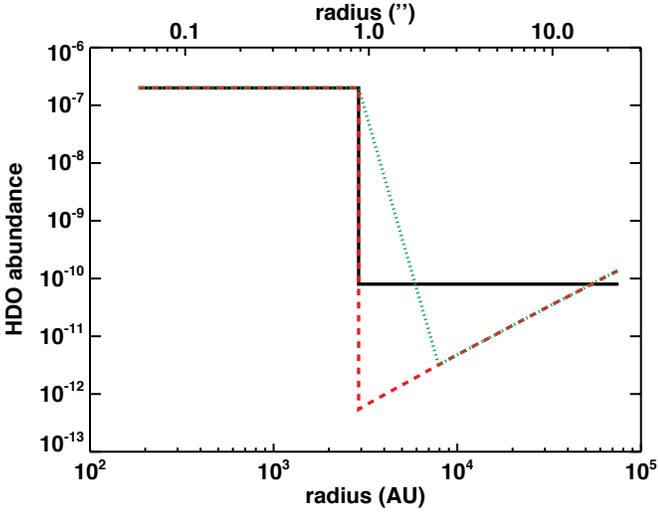}
\caption{Best-fit abundance profiles obtained for HDO when the outer abundance (region at $T$ $<$ 200 K) is constant (black solid line, Section \ref{sect_jump}), when it decreases from the cold regions to the region at $T$ = 200 K (red dashed line, Section \ref{sect_slope}), and when it decreases from the cold regions to the region at $T$ = 100 K and increases from $T$ = 100\,K to $T$ = 200\,K (green dotted line, see Section \ref{sect_mod_increase}). The temperature reaches 200\,K at a radius of 2700 AU ($\sim$0.9$\arcsec$, see Figure B1). (A color version of this figure is available in the online journal.)}
\label{water_profiles}
\end{center}
\end{figure}

Although we used a constant abundance of HDO and H$_2^{18}$O in the cold envelope of G34, it is very probable that the water abundance shows variations in this region due to non-thermal desorption mechanisms.
In particular, \citet{Mottram2013} showed that the desorption by the cosmic ray-induced UV field leads to an outer abundance of water decreasing gradually from the cold to the warm regions of low-mass protostars.
To confirm that the presence of a gradual abundance decrease in the cold envelope does not affect the derived value of the HDO/H$_2$O ratio in this region, we ran a modeling considering an equilibrium state between the desorption by the cosmic ray-induced UV field and the re-depletion on the grains. Using similar equations to those in \citet{Hollenbach2009} and \citet{Mottram2013}, we get by equating desorption to depletion: 
\begin{equation}
G_{\rm cr} F_{\rm 0} Y_{\rm x} f_{\rm s,x} n_{\rm gr} \sigma_{\rm gr} = n{\rm (x)} n_{\rm gr} \sigma_{\rm gr} \varv_{\rm th,x}
\end{equation}
with $F_{\rm 0}$ the local interstellar flux of 6--13.6 eV photons assumed to be equal to 10$^8$ photons cm$^{-2}$ s$^{-1}$, $G_{\rm cr}$ the scaling factor of the UV flux, $Y_{\rm x}$ the photodesorption yield for the molecule x ($\sim$\,10$^{-3}$ for H$_2$O, \citealt{Oberg2009}), $f_{\rm s,x}$ the fraction of the molecule x on grains, $n_{\rm gr}$ the grain density, $\sigma_{\rm gr}$ the cross sectional area of the grain and $\varv_{\rm th,x}$ the thermal velocity.
The thermal velocity is calculated according to the following formalism:
\begin{equation}
\varv_{\rm th,x} = \sqrt{  \frac{8 k_{\rm b} T_{\rm k}}{\pi m_{\rm x}}},
\end{equation}
where $k_{\rm b}$ is the Boltzmann constant, $T_{\rm k}$ the gas temperature and $m_{\rm x}$ the mass of the molecule x.
The outer abundance of H$_2$O with respect to H$_2$ is then equal to:
\begin{equation}
 X_{\rm out}{\rm (H_2O)} = \frac{G_{\rm cr} F_{\rm 0} Y_{\rm H_2O} f_{\rm s,H_2O}}{\varv_{\rm th,H_2O}~n_{\rm H_2}},
\end{equation}
with $n_{\rm H_2}$ the H$_2$ density.
Similarly we obtain for HDO:
 \begin{equation}
 X_{\rm out}{\rm (HDO)} = \frac{G_{\rm cr} F_{\rm 0} Y_{\rm HDO} f_{\rm s,H_2O}}{\varv_{\rm th,HDO}~n_{\rm H_2}} \left(\frac{\rm HDO}{\rm H_2O}\right),
\end{equation}
and for H$_2^{18}$O:
 \begin{equation}
 X_{\rm out}{\rm (H_2^{18}O)} = \frac{G_{\rm cr} F_{\rm 0} Y_{\rm H_2^{18}O} f_{\rm s,H_2O}}{\varv_{\rm th,H_2^{18}O}~n_{\rm H_2}} \left(\frac{\rm H_2^{18}O}{\rm H_2O}\right).
\end{equation}
The photodesorption yields for HDO and H$_2^{18}$O are assumed similar to those for H$_2$O \citep{Oberg2009}.
The thermal velocity is approximatively the same due to their relatively similar masses.
All the other parameters are independent of the molecules except the fraction f$_{\rm s,x}$ of these molecules contained in the grain mantles which reflects the isotopic ratios, HDO/H$_2$O and H$_2^{18}$O/H$_2^{16}$O, on the grains. The external UV field should also affect the external part of the outer envelope. But, due to the very small constraints on these different mechanisms, we only considered the desorption by the cosmic ray-induced UV field.

\begin{figure*}
\begin{center}
\includegraphics[width=1.0\textwidth]{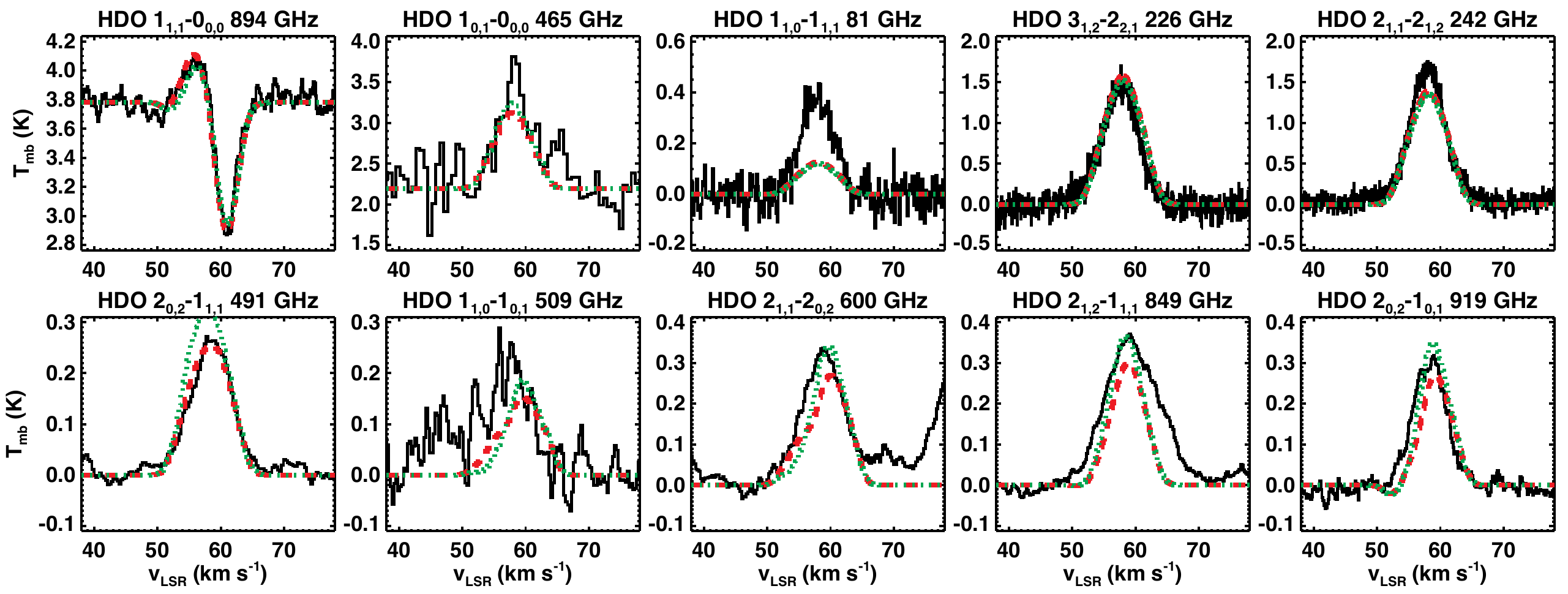}
\caption{{\it Black solid line}: HDO lines observed with HIFI, IRAM, and CSO.
{\it Red dashed line}: Modeling for a constant inner abundance $X_{\rm in}$ = 2\,$\times$\,10$^{-7}$  (T $\geq$ 200\,K) and an outer abundance (T $<$ 200\,K) decreasing from the cold to the warm regions (see Section \ref{sect_slope}).
{\it Green dotted line}: Modeling for a constant inner abundance $X_{\rm in}$ = 2\,$\times$\,10$^{-7}$  (T $\geq$ 200\,K), an abundance gradually increasing from 100 to 200\,K, and an outer abundance (T $<$ 100\,K) decreasing from the cold to the warm regions (see Section \ref{sect_mod_increase}). (A color version of this figure is available in the online journal.)}
\label{ratran_hdo_Gcr}
\end{center}
\end{figure*}

\begin{figure*}
\begin{center}
\includegraphics[width=0.62\textwidth]{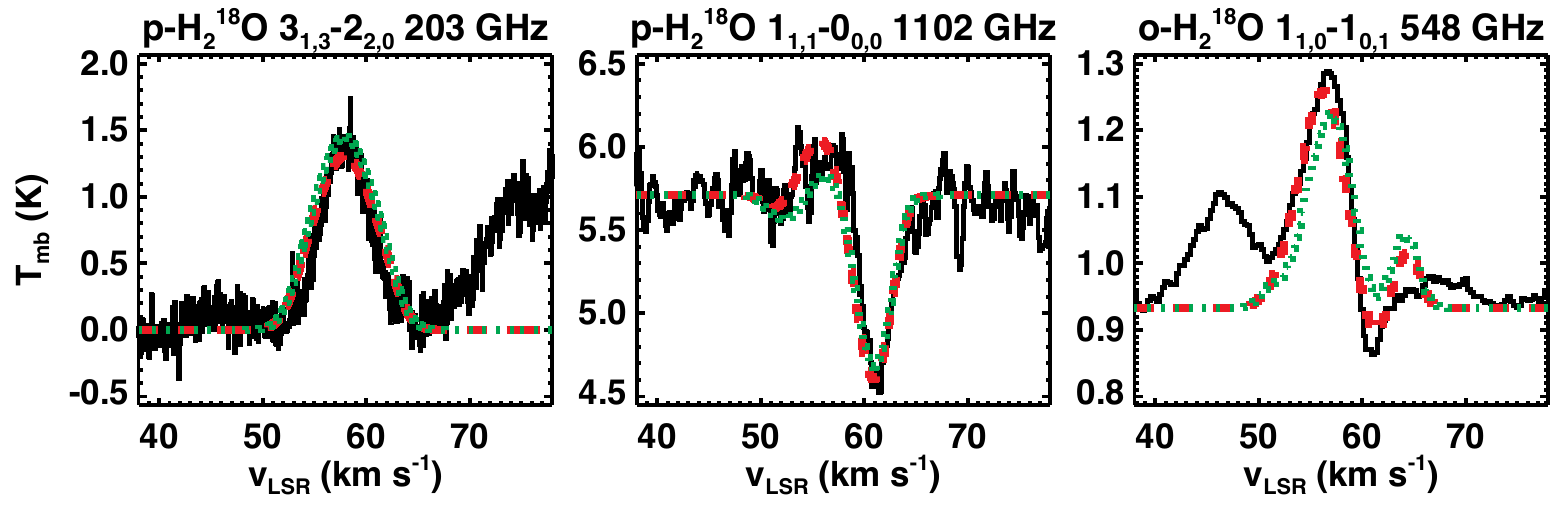}
\caption{{\it Black solid line}: H$_2^{18}$O lines observed with HIFI and IRAM.
{\it Red dashed line}: Modeling for a constant inner abundance $X_{\rm in}$ = 9\,$\times$\,10$^{-7}$  (T $\geq$ 200\,K) and an outer abundance (T $<$ 200\,K) decreasing from the cold to the warm regions (see Section \ref{sect_slope}).
{\it Green dotted line}: Modeling for a constant inner abundance $X_{\rm in}$ = 9\,$\times$\,10$^{-7}$  (T $\geq$ 200\,K), an abundance gradually increasing from 100 to 200\,K, and an outer abundance (T $<$ 100\,K) decreasing from the cold to the warm regions (see Section \ref{sect_mod_increase}). (A color version of this figure is available in the online journal.)}
\label{ratran_h218o_Gcr}
\end{center}
\end{figure*}

We ran a grid of models for the case $T_{\rm j}$ = 200\,K, keeping the inner abundances determined previously  ($X_{\rm in}$(HDO) = 2\,$\times$\,10$^{-7}$ and $X_{\rm in}$(H$_2^{18}$O) = 9\,$\times$\,10$^{-7}$). Different values were then assumed for the factors $W_{\rm HDO} = G_{\rm cr} f_{\rm s,H_2O}~{\rm HDO/H_2O}$ and $W_{\rm H_2^{18}O} = G_{\rm cr} f_{\rm s,H_2O} {\rm ~H_2^{18}O/H_2^{16}O}$. Assuming H$_2^{18}$O/H$_2^{16}$O = 400 \citep{Wilson1999} and $f_{\rm s,H_2O}$ = 1 (the icy grain mantles are constituted entirely of H$_2$O), the best-fit model of the H$_2^{18}$O lines gives a scaling factor $G_{\rm cr}$ of about 1.6\,$\times$\,10$^{-3}$. If water represents only 50\% of the grain mantles, $G_{\rm cr}$ is then equal to 3.2\,$\times$\,10$^{-3}$ leading to a cosmic ray-induced UV field of $\sim$3\,$\times$\,10$^5$ photons cm$^{-2}$ s$^{-1}$. 
These values represent, however, only upper limits, since the desorption by the external UV field is not taken into account in the analysis. The typical value of the cosmic-ray induced UV flux \citep[$G_{\rm cr}$ $\sim$ 10$^{-4}$; e.g.,][]{Prasad1983,Shen2004} is then consistent with the upper limit derived here ($G_{\rm cr}$ $\lesssim$ 3\,$\times$\,10$^{-3}$).

 The best-fit abundance profile determined for HDO when the outer abundance decreases from the cold to the warm regions is presented in Figure \ref{water_profiles} (red dashed line). The HDO/H$_2$O ratio in the outer envelope is equal to 1.3\,$\times$\,10$^{-3}$. It is then, once again, higher than in the hot core ($\sim$(5--6)\,$\times$\,10$^{-4}$). 
The HDO and H$_2^{18}$O line profiles predicted with the RATRAN code (see red dashed lines in Figures \ref{ratran_hdo_Gcr} and \ref{ratran_h218o_Gcr}) are relatively similar to those in Figures \ref{ratran_hdo_200K} and \ref{ratran_h218o_200K} that assume a two-step abundance profile with a jump at 200\,K.  The fit is even better for the H$_2^{18}$O and HDO fundamental transitions (HDO: 894 GHz; H$_2^{18}$O: 548 and 1102 GHz), as the predicted intensity of their emission is now in agreement with the observations. Some of the HDO lines (509, 600, and 919 GHz) show small self-absorptions on their blue-shifted side. However, these defects could probably disappear with slightly different velocity profiles. Indeed, the velocity profiles used here were only adapted for the models with the abundance jumps (see Appendix B2).

\subsubsection{Gradual increase of the water abundance profile at the cold envelope/hot core transition}
\label{sect_mod_increase}

In Section \ref{sect_jump}, we mentioned that a model with a gradual increase of the HDO abundance at the cold envelope/hot core transition could potentially explain why we need a higher jump temperature than 100\,K to reproduce the HDO line profiles.
Here we show the results obtained with both a decrease of the outer abundance from the outermost regions to the regions at 100 K and a gradual increase from 100 to 200\,K. This type of profile is then relatively similar to the predictions of chemical models coupled with a dynamical approach \citep[][Wakelam et al. submitted]{Aikawa2012}. The HDO inner abundance is equal to 2\,$\times$\,10$^{-7}$ and the outer abundance follows the trend described in Section \ref{sect_slope}.
The abundance profile used and the result of the model for HDO are shown in Figures~\ref{water_profiles} and \ref{ratran_hdo_Gcr} (green dotted line), respectively. 
This model also appears very similar to the model with an abundance jump at 200 K (see Figure \ref{ratran_hdo_200K}).  
A model with both an abundance decrease (with temperature) in the colder envelope and an increase of the abundance towards the hot core is probably more realistic than the jump abundance assumption and could explain why the hot core is smaller than expected. It is, however, important to note that the temperature range of the gradual abundance increase is not known. We assume here the range 100--200\,K but it could be slightly different and a specific range is probably dependent on the dynamics. The result of this modeling should thus be considered only qualitatively. 
We can however conclude that this type of abundance profile allows to reproduce the HDO line profiles as well as the abundance jump models at $\sim$150--220\,K.

We ran a similar model for the H$_2^{18}$O lines. 
The model (presented in Section \ref{sect_mod_increase}) with a gradual increase of the abundance at the cold envelope/hot core transition is presented in Figure \ref{ratran_h218o_Gcr} (green dashed lines). The lines are here again reproduced as well as by the jump abundance models. The HDO/H$_2$O ratio shows consequently the same variation between the inner and outer regions as found before, i.e. 5.6\,$\times$\,10$^{-4}$ at $T$ $>$ 200\,K and 1.3\,$\times$\,10$^{-3}$ at $T$~$<$~100\,K.

\subsection{Comparison with previous studies}

The singly deuterated form of water has been studied toward many high-mass hot cores with ground-based telescopes \citep{Jacq1990,Gensheimer1996,Pardo2001,vandertak2006}. These studies are relevant for the hot core study but do not directly address for the cold external envelope, since the observations of the ground HDO transition at 894 GHz with a very good signal-to-noise ratio are necessary in order to disentangle the contribution from the hot core to the contribution of the cold envelope. The launch of the \textit{Herschel} Space Observatory dramatically changed the situation, with the access to the high frequency range with many HDO transitions available in addition to the ground-state transition. The D/H ratio in water remained for a long time very poorly known since the study of water was based on observations suffering from dilution in the large beams of the Infrared Space Observatory (ISO), the Submillimeter Wave Astronomy Satellite (SWAS) and the ODIN satellite as well as from large opacities. The only way to study water from the ground was to use the H$_2^{18}$O transition available with some telescopes at 203 GHz \citep{Jacq1988,vandertak2006,Jorgensen2010b,Persson2012,Persson2013}. 
With the help of this line, the water deuterium fractionation was previously estimated in the high-mass star-forming region G34 by \citet{Jacq1990}, \citet{Gensheimer1996}, and \citet{Liu2013}. 
They found, in its hot core, HDO/H$_2$O ratios ranging between 1\,$\times$\,10$^{-4}$ and 4\,$\times$\,10$^{-4}$. Since our modeling in the hot core region is mostly dominated by the 81, 226 and 241 GHz transitions accessible from the ground, these values are relatively consistent with our estimate of (5--6)\,$\times$\,10$^{-4}$ both with the rotational diagram approach and the non-LTE 1D analysis.
Note that we assumed an H$_2$$^{16}$O/H$_2$$^{18}$O ratio of 400, whereas the previous studies assumed 500.
In addition, the HDO/H$_2$O ratio found in this hot core is consistent with the average HDO/H$_2$O ratio (a few 10$^{-4}$) found in other high-mass sources \citep{Jacq1990,Gensheimer1996,Pardo2001,vandertak2006,Emprechtinger2013}.

In the hot core, we also determined the water abundance (relative to H$_2$) to be a few\,$\times$\,10$^{-4}$. Similar values were estimated in other high-mass hot cores \citep{Chavarria2010,Herpin2012,Neill2013}, although lower values were also found, for example, in NGC~6334~I ($\sim$10$^{-6}$, \citealt{Emprechtinger2013}). The value of 10$^{-4}$ is comparable to the observed abundance of solid water and together with the derived HDO/H$_2$O abundance ratios of $10^{-4}-10^{-3}$ suggests that the origin of the observed water is evaporation of grain mantles.  

Recently, \citet{Liu2013} also attempted to constrain the D/H ratio for water in the outer envelope of G34 using the 894 GHz transition observed from the ground with APEX. 
From a RATRAN modeling using an abundance jump profile at 100\,K, they failed to reproduce the profile of this ground state transition leading to a very uncertain value for the D/H ratio in the outer region of the envelope of ($1.9$--$4.9$)\,$\times$\,10$^{-4}$.
With the sensitivity of \textit{Herschel}/HIFI observations of the 894 GHz transition, it became possible to measure accurately the D/H ratio of water in low-mass \citep{Coutens2012,Coutens2013b} and high-mass protostars, from the hot core region to the cold external envelope. 
We showed here that, with a value of (1.0--2.2)\,$\times$\,10$^{-3}$ in the colder envelope, the HDO/H$_2$O ratio is indeed higher than the estimate by \citet{Liu2013}. It is also higher than in the hot core.
A similar behavior was discovered in the low-mass sources IRAS16293 and NGC1333 IRAS4A (Coutens et al. 2013a, 2013b). 
But this is the first time that a radial variation of the D/H ratio has been observed towards a high-mass star-forming region. 
The HDO/H$_2$O ratio derived in the colder envelope of G34 is among the highest values found in high-mass sources. It is close to the high value of (2--4)\,$\times$\,10$^{-3}$ found in Orion KL \citep{Persson2007,Neill2013} but lower by more than a factor 10 than in the absorbing layer of low-mass protostars \citep{Coutens2012,Coutens2013b}.

\section{Chemical modeling}
\label{sect_chemistry}

In order to study the chemical pathways that could lead to the observed HDO and H$_2$O abundances and their corresponding ratio, we modeled the chemical evolution of the source as a function of its radius, using the full gas-grain chemical model Nautilus \citep{Hersant2009}. 

\subsection{Model}
\label{subsubsec:model}

Nautilus is a gas grain chemical code adapted from the original code developed by the Herbst group \citep{Hasegawa1993}.
It solves the kinetic equations of gas-phase chemistry, takes into account grain surface chemistry, and interactions between both phases (adsorption, thermal and non-thermal desorption).
The rate equations follow \citet{Hasegawa1992} and \citet{Caselli1998}.
More details on the processes included in the code are presented by \citet{Semenov2010}.
The chemical network is adapted from \citet{Aikawa2012} and \citet{Furuya2012}.
As pointed out by \citet{Pagani1992}, \citet{Flower2004,Flower2006a,Flower2006b}, \citet{Walmsley2004}, and \citet{Pagani2009}, considering ortho and para spin modifications of various H and D bearing species is important due to some reactions which are much faster with ortho--H$_2$ than para--H$_2$, and can change the entire chemistry of deuterium fractionation.
Thus, we extended the network including the ortho, para, and meta states of H$_2$, D$_2$, H$_3^+$, H$_2$D$^+$, D$_2$H$^+$, and D$_3^+$.
For the reactions involving these species, we have applied spin selection rules to know which reactions are allowed, and have determined branching ratios assuming a total scrambling and a pure nuclear spin statistical weight.
Some of the rate coefficients of these reactions have been theoretically or experimentally determined \citep{Marquette1988,Jensen2000,McCall2004,DosSantos2007,Hugo2009,Honvault2011a,Honvault2011b,Dislaire2012} and for these we used the calculated or measured values.
We have benchmarked our model against some previous work that includes spin-state chemistry, using the same conditions as described in Figure~8 of \citet{Pagani2009} and Figure~4 of \cite{Sipila2013}: a temperature of $\sim$10~K and a density from $\sim$10$^5$ to $\sim$10$^6$~cm$^{-3}$.
Minor differences in abundances do exist, since the networks, the models, and the input parameters can be slightly different, but the result is globally similar.
A notable difference is however seen for HD after $10^5$~yrs as compared with \cite{Sipila2013}.
They predict a decrease of its gas phase abundance by one order of magnitude at $10^6$~yr.
Under the same conditions, we predict a decrease in the gas phase HD abundance of only a factor $\approx 2$, similar to the model of \citet[][priv. com.]{Albertsson2013,Albertsson2014}.
The inclusion in our model of photodesorption and reactive desorption may have some effect on HD depletion.
Photodesorption due to direct interstellar UV photons and secondary photons generated by cosmic rays, as well as the exothermic association between the surface species H and D, may both release enough HD molecules to the gas phase to lower the HD depletion.
The network and a benchmark will be presented in more detail in a forthcoming paper (U. Hincelin et al., in preparation).

In our model, elemental and initial abundances follow \citet{Hincelin2011}.
Initially, the ortho-to-para H$_2$ ratio is set to its statistical value of 3, and deuterium is assumed to be entirely in HD form with an abundance of $1.5 \times 10^{-5}$ relative to total hydrogen, following \citet{Kong2013}.
Note that the timescale for conversion to a thermal ortho-to-para H$_2$ ratio is a few times $10^5$ to a few times $10^6$~yr at 10~K depending on the density, as in \cite{Pagani2009}.
In the evolutionary sequence of high-mass star formation proposed by \citet{Beuther2007} and \citet{Zinnecker2007}, infrared dark clouds (IRDCs) are expected to be the first stage.
Comparing observations of high-mass star-forming regions with advanced gas-grain chemical modeling, \cite{Gerner2014} derived a chemical age for this stage of around $10^4$~yrs.
The mean density and temperature of IRDCs are respectively $10^5$~cm$^{-3}$ and 16~K \citep{Sridharan2005}.
From the initial elemental and chemical abundances, we have computed the chemical evolution over a period of  $10^4$~yrs, corresponding to $t_{IRDC}$ in Figure~\ref{fig_chemistry_HDO_H2O},  with a temperature of 16~K, a proton density of $2 \times 10^{5}$~cm$^{-3}$, and a visual extinction of 30.  In our standard model, we use a cosmic ray ionization rate of $1.3 \times 10^{-17}$~s$^{-1}$, but also use a value ten times higher, as discussed in Section \ref{sect_chem_results}. 
Following this first phase, we switched to a time-independent one-dimensional physical structure of G34 derived by \citet{vanderTak2013} as seen in Figure B1, and allowed the time-dependent chemistry to continue to evolve independently at each value of the radius of the source.

\subsection{Results}
\label{sect_chem_results}

\begin{figure}
\includegraphics[width=1.0\linewidth]{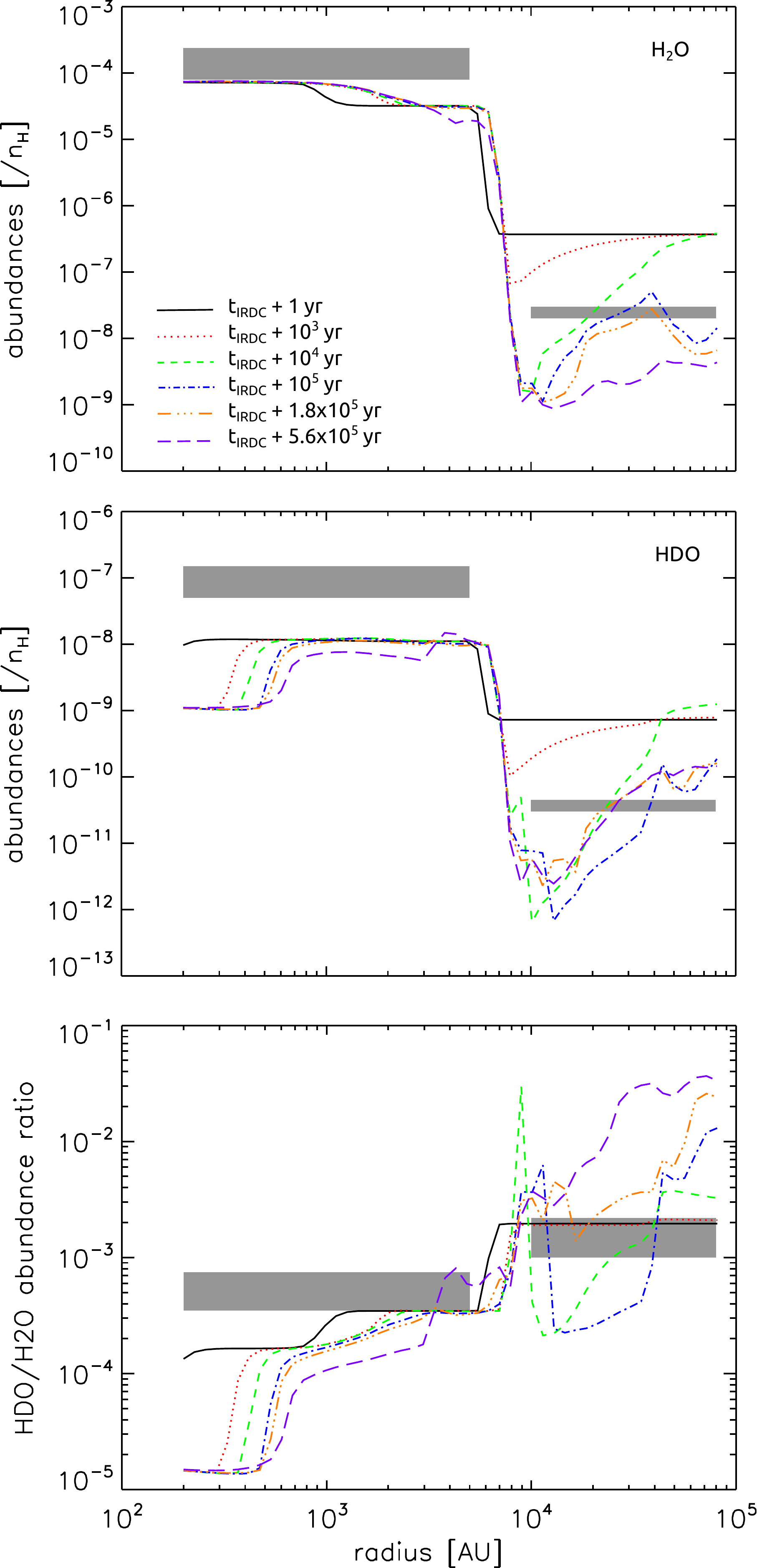}
\caption{Top and center: calculated gas-phase abundances of H$_2$O and HDO relative to the total density of protons.
Gray areas show observational values and uncertainties of H$_2$O and HDO, observed in the hot inner core and the colder envelope.
Bottom: HDO/H$_2$O gas-phase abundance ratio.
Gray areas show observational values and uncertainties of HDO/H$_2$O gas-phase abundance ratio observed in the hot inner core and the colder envelope (see Sections \ref{sect_jump} and \ref{sect_h218o} for information about uncertainties).
Both abundances are plotted as a function of the radius of the source.
The results are time dependent, and the colors and types of lines correspond to different values after the first initial phase: black solid lines ($t=t_{IRDC}+1$~yr), red dotted lines ($t=t_{IRDC}+10^3$~yr), green short dashed lines ($t=t_{IRDC}+10^4$~yr), blue dashed dotted (1 dot) lines ($t=t_{IRDC}+10^5$~yr), orange dashed dotted (3 dots) lines ($t=t_{IRDC}+1.8\times10^5$~yr), and purple long dashed lines ($t=t_{IRDC}+5.6\times10^5$~yr). (A color version of this figure is available in the online journal.)}
\label{fig_chemistry_HDO_H2O}
\end{figure}

Figure~\ref{fig_chemistry_HDO_H2O} shows the computed fractional abundances for  gaseous HDO and H$_2$O  relative to the total proton density and their ratio as a function of the radius of the source, at different times following the IRDC stage.
The computed values can be compared with the values that best fit the observations, as listed in Table \ref{table_results}.
The observational values are given for two points in the table, the inner hot core and the colder envelope, but these values are represented as areas in the figures with their height referring to uncertainty and their length to the length of the inner and outer regions. 
Note that observational results may not be constant as a function of radius, as shown for the abundances in Sections \ref{sect_slope} and \ref{sect_mod_increase}.

During the IRDC phase, water and HDO are present mainly on the grain surfaces, with the water abundance $\approx 10^{-4}$.
Once we apply the physical profile of the source, the temperature in the inner region, greater than $\sim$100~K, is high enough to allow the rapid desorption of H$_2$O and HDO, and a transition region is observed around $6\times 10^3$~AU, which corresponds to $\sim$100~K.
Beyond $6\times 10^3$~AU, the reverse effect is observed: molecules are slowly adsorbed onto grain surfaces depending on the radius, because the density of the source is now higher than during the IRDC phase.
The rate of adsorption is directly proportional to the density, and since the density is higher for small radii, the gaseous molecules are adsorbed more quickly closer to the transition region.
This effect is clearly seen at times of 10$^{3}$ yrs and longer. 
While the  gas-phase water fractional abundance predicted by the chemical model in the inner core (radius $\leq 5000$~AU) is almost constant, at $10^{-4}$ to $10^{-5}$ relative to the total proton density,
in the colder envelope, the water abundances lie between a few\,$\times$\,$10^{-7}$ and $10^{-9}$ depending on the radius and the time.
This dependence also holds for HDO, which possesses an inner-core abundance between $10^{-8}$ and $10^{-9}$, and an outer abundance between a few\,$\times$\,$10^{-9}$ and $10^{-12}$.

In addition to these gas-grain interactions, chemical reactions are also occurring.
In the inner core, gaseous water is mainly destroyed by  reactions with atomic hydrogen: $\rm H + H_2O \longrightarrow OH + \textit{ortho-}H_2$, and $\rm H + H_2O \longrightarrow OH + \textit{para-}H_2$.
However, water is efficiently reformed by the reverse reactions, so its abundance does not change significantly.
In the same region, HDO is also mainly destroyed by reactions with atomic hydrogen: $\rm H + HDO \longrightarrow OH + HD$, $\rm H + HDO \longrightarrow OD + \textit{ortho-}H_2$, and $\rm H + HDO \longrightarrow OD + \textit{para-}H_2$.
Although HDO is also reformed by the reverse reactions, these processes are sufficiently slower than the destruction reactions that the HDO abundance decreases, with an efficiency depending on the local temperature.
This is indicated by the dashed lines in the upper panel of Figure~\ref{fig_chemistry_HDO_H2O}, particularly within a radius of 1000~AU.
Thus, we observe a general decrease of the HDO/H$_2$O ratio in the hot inner core as a function of time.  

In the colder envelope, at larger radii, the H$_2$O gas phase abundance is reduced due to adsorption, as discussed above, and ion-molecule reactions, particularly the reaction with HCO$^+$, which forms H$_3$O$^+$ and CO.
Before $t=t_{IRDC}+10^4$~yrs, HCO$^+$ mainly reacts with carbon atoms, and after this time, the carbon atom abundance is low enough to allow an increase of the HCO$^+$ abundance through ion-molecule reactions involving CO.
Although HDO also reacts with HCO$^+$, it is partially reformed by ion-molecule reactions involving $\rm H_2DO^+$, and dissociative recombination of $\rm H_2DO^+$ with an electron.
H$_2$O is also reformed by reactions involving H$_3$O$^+$, but not as efficiently as HDO.
At later times, the abundances of HDO$^+$ and H$_2$DO$^+$ are increased, while the ones of H$_2$O$^+$ and H$_3$O$^+$ are decreased, so that the HDO/H$_2$O abundance ratio increases.

At $10^4$~AU, next to the transition region, the temperature and density are respectively equal to 80~K and $10^6$~cm$^{-3}$.
Here, there is a complex competition between the formation of HDO and H$_2$O in the gas phase and the adsorption and desorption of these molecules.
For this reason, we get temporarily a peak in the HDO/H$_2$O ratio around $10^4$ and $10^5$ yrs (respectively the green and blue peak).
The main gas phase reactions involved are the following: H$_3$O$^+$ and H$_2$DO$^+$ react with DCN, DNC, HCN, and HNC, which form HDO and H$_2$O.
Besides, after $10^4$~yrs, H$_2$CO plays also a role: it is slowly released from the grain surface, and reacts efficiently with OH and OD to form respectively H$_2$O and HDO.
However, at this temperature and density, adsorption of HDO and H$_2$O is still quite efficient, and removes a part of these molecules from the gas phase.

If we compare the computed abundances of water and HDO with the observational values, seen as gray areas in Figure~\ref{fig_chemistry_HDO_H2O}, the H$_2$O abundances are in good agreement in both the hot inner core and the colder outer  envelope.
This also holds true for HDO in the colder envelope; however, our model does not produce enough HDO in the hot inner core at all times.
Specifically, our values are five to fifty times less than those indicated by the observations, depending on the time and the radius.
Given the low abundance of HDO in the hot inner core, our calculated gaseous HDO/H$_2$O ratio is lower than the observed one throughout this region, while in the colder outer envelope, our ratio lies within the range of the observational values at selected times.
Note that the observational abundances and ratio may not be constant as a function of radius in the two regions, so more constraints are necessary to compare with the model results.

The HDO abundance profiles in the cold outer region from Figure~\ref{fig_chemistry_HDO_H2O} favor the best-fit abundance profile for water from Figure \ref{water_profiles}, which increases with radius in the cold envelope.
A comparison between the HDO profiles of these two figures leads to the best agreement around $t=t_{IRDC}+10^5$~yrs.
This time corresponds to the best-fit chemical age of \citet{Gerner2014}: their high mass protostellar object stage, the stage just after the IRDC stage, lasts $\sim 6\times 10^4$~yrs, and the following stage, the hot molecular core stage, lasts $\sim 4\times 10^4$~yrs, which give a total similar to ours. 

We have tested the sensitivity of H$_2$O and HDO to the cosmic-ray ionization rate, using a value of $1.3 \times 10^{-16}$~s$^{-1}$, which is ten times higher than the standard rate.
This value is close to the upper limit derived in this source (see Section~\ref{sect_slope}).
Cosmic rays are the main source of ions in clouds, and formation and destruction of neutral species involve mainly reactions with charged species.
As a consequence, most of the molecules are sensitive to the cosmic-ray ionization rate \citep{Wakelam2010}.
Compared with our standard model, in the cold envelope, the gas phase H$_2$O abundance is decreased by one order of magnitude at early times after the IRDC phase. 
Then, H$_2$O is reformed quite efficiently so that the final abundance is one order of magnitude higher than with our standard model.
In the same region, the HDO abundance is increased by a factor 10 to 100 depending on the time.
The HDO/H$_2$O ratio is then enhanced, and higher than the observational value by a factor of $\sim$100 and $<$2 respectively at early and later times.
In the inner core, the H$_2$O abundance is slightly increased to a value $\geq 10^{-4}$ at all times.
The HDO abundance is more sensitive at early times to the cosmic ionization rate: it is firstly increased by a factor of 100, but then the value tends to decrease to the same one as in our standard model. 
The HDO/H$_2$O ratio is also enhanced, up to a factor of 100 at early times, but tends to decrease to the same value as in our standard model. 
In the IRDC phase and the cold envelope, gaseous H$_2$O is mainly formed by reactions involving H$_3$O$^+$ and destroyed by reactions with HCO$^+$ and C$^+$, while
H$_2$DO$^+$ is the main reactant involved in the formation of HDO.
In the inner hot core, the abundances of H$_2$O and HDO are mainly changed due to OH and OD which are sensitive to the cosmic ray ionization rate \citep{Wakelam2010}.

We have also tested the sensitivity of our modeling to the inclusion of spin-state chemistry, and provide in Appendix C the results of a simulation using our chemical network without considering the spin states.
Our main conclusion is that the gas phase HDO abundance is not only sensitive to the inclusion of spin-state chemistry at low temperature, but also at high temperature, although the difference is less strong.
In addition, the H$_2$O abundance is slightly sensitive at longer times to the spin-state chemistry in the cold envelope region, but not in the hot inner core region.  
The overall ratio HDO/H$_2$O decreases if we take into account spin state chemistry, as it can be predicted based simply on the thermodynamics of protonated ion-HD exchange reactions.

\subsection{Comparison with other studies}

Below we compare our results for water and HDO both in the gas and on ice mantles with those of earlier studies.
We first consider ice mantles.
Some of these studies include spin-state chemistry while others do not.

Several groups theoretically studied deuteration of water in star forming regions \citep[i.e.][]{Cazaux2011,Aikawa2012,Sipila2013,Taquet2013,Taquet2014}. 
These studies focus on low mass star-formation regions or cold conditions, and as a consequence generally deal with lower temperatures and densities than ours.
However, considering the external region of the cold envelope of our source, where the conditions are the closest to these studies (30~K and $10^5$~cm$^{-3}$), it is worth making some comparisons with our ice results.
Our HDO/H$_2$O ice  ratio in the cold envelope varies between $10^{-4}$ and $10^{-3}$ depending on the time.
The larger the time, the larger  the ratio.
We can compare our values to those in Figures~11 and 12 in \cite{Sipila2013} and Figure~8 in \cite{Taquet2013}.
These studies also include spin state chemistry.
In general, we predict a lower HDO/H$_2$O ice ratio than these studies.
Despite our slightly higher temperature, and multiple differences between our models, the initial ortho-to-para H$_2$ ratio may be the main reason, since a higher value tends to decrease the deuterium fractionation.
\cite{Cazaux2011} and \cite{Aikawa2012}, who did not consider the spin state chemistry, predicted an HDO/H$_2$O ice ratio of $\sim 0.01$, which then can be considered as an upper limit.

\citet{Aikawa2012} and \citet{Taquet2014} studied the deuteration of molecules as a function of the radius of a forming protostellar core. Here we can compare calculated HDO/H$_{2}$O ratios in the gas phase.
Their temperature and density gradient along the radius is quite important, from $\sim10$~K and $\sim10^4$~cm$^{-3}$ to several hundred Kelvin and $10^{12}$~cm$^{-3}$, close to the range of conditions of our source.
Note that these studies include a dynamical physical structure instead of a static structure.
Despite the differences between our model and these earlier studies, we obtain the same qualitative pattern, in which the
gas phase water abundance is higher in the inner and hot region, while it is lower in the outer and cold region.
In the outer region, the abundance is governed mainly by the density, and as a consequence, tends to be lower when the density gets higher.
Their HDO/H$_2$O ratio changes by one to two orders of magnitude between the cold region and the hot region, and is higher in the colder region.

\section{Conclusions}
\label{sect_conclu}

Ten lines of HDO and three lines of H$_2^{18}$O covering a broad range of upper energy levels (22--204\,K) were detected with the \textit{Herschel}/HIFI instrument, the IRAM-30m telescope, and the CSO towards the high-mass star-forming region G34.26+0.15. 
Using a 1D non-LTE radiative transfer code, we constrained the abundance distribution of HDO and H$_2^{18}$O throughout the envelope from the hot core to the colder regions.
To reproduce the HDO line profiles, it is necessary to assume an abundance jump at a temperature higher than 100\,K ($\sim$150--220\,K), which suggests that the hot core is smaller than expected. This could be explained by the fact that the water abundance increases gradually within a certain temperature range between the cold envelope and the hot core, as suggested by some chemical models considering dynamics \citep[][Wakelam et al. submitted]{Aikawa2012}. 
Another explanation would be that the structure is relatively uncertain at small scales. Similar studies (including observations of the HDO lines at 226 and 242 GHz, as well as some HIFI lines at 491, 600, or 919 GHz) should be carried out towards other high-mass sources to know if this higher jump temperature is specific to G34 or common to other objects. 
Using different types of water abundance profiles, we showed that the water deuterium fractionation in the hot core and in the colder envelope is strongly constrained.
Assuming calibration uncertainties of 20\%, the HDO/H$_2$O ratio is estimated to be about (3.5--7.5)\,$\times$\,10$^{-4}$ in the hot core. It is in agreement with the value derived with the rotational diagram analysis of the IRAM-30m lines as well as with previous studies \citep{Jacq1990,Liu2013}. In the colder gas, we determined the HDO/H$_2$O ratio to be about (1.0--2.2)\,$\times$\,10$^{-3}$, including the uncertainties. Although radial variations of the water deuterium fractionation have already been observed in low-mass protostars (Coutens et al. 2012, 2013a, 2013b), this is the first time that 
a decrease of the water deuterium fractionation in the warmer regions has been measured in a high-mass star-forming region. 
Finally, we modeled the chemical evolution of G34 as a function of its radius and showed that our model reproduces relatively well the observational results that assumed an increase of the water abundance with radius in the cold regions (see Figures \ref{water_profiles} and \ref{fig_chemistry_HDO_H2O}).
The comparison of the chemical model and the observations favors an age of 10$^5$ years after the IRDC stage, which is consistent with the age derived for hot molecular cores by \citet{Gerner2014}.

\section*{Acknowledgments}
 The authors are grateful to the anonymous referee for his/her useful and pertinent comments and suggestions. 
  They thank K. Furuya and Y. Aikawa for providing the initial chemical network of deuterated species and N. Flagey for providing the reduced HIFI data of H$_2^{18}$O. They would also like to thank M. Hajigholi for fruitful discussions regarding the source modeling. A.\,C. and C.\,V.  thank PCMI for support of the Herschel HIFI project on deuterated water.
  C.\,M.\,P. acknowledges generous support from the Swedish National Space Board. Support for this work was also provided by NASA through an award issued by JPL/Caltech.
  
  This work is based on observations carried out with the HIFI instrument onboard the \textit{Herschel Space Observatory}, the Institut de RadioAstronomie Millim\'etrique (IRAM) 30m Telescope and the Caltech Submillimeter Telescope (CSO). {\it Herschel} is an ESA space observatory with science instruments provided by European-led principal Investigator consortia and with important participation from NASA. 
  HIFI has been designed and built by a consortium of institutes and
  university departments from across Europe, Canada and the United
  States under the leadership of SRON Netherlands Institute for Space
  Research, Groningen, The Netherlands and with major contributions
  from Germany, France and the US. Consortium members are: Canada:
  CSA, U.Waterloo; France: IRAP (formerly CESR), LAB, LERMA, IRAM; Germany: KOSMA,
  MPIfR, MPS; Ireland, NUI Maynooth; Italy: ASI, IFSI-INAF,
  Osservatorio Astrofisico di Arcetri-INAF; Netherlands: SRON, TUD;
  Poland: CAMK, CBK; Spain: Observatorio Astron\'omico Nacional (IGN),
  Centro de Astrobiolog\'{\i}a (CSIC-INTA). Sweden: Chalmers
  University of Technology - MC2, RSS \& GARD; Onsala Space
  Observatory; Swedish National Space Board, Stockholm University -
  Stockholm Observatory; Switzerland: ETH Zurich, FHNW; USA: Caltech,
  JPL, NHSC.  
IRAM is supported by INSU/CNRS (France), MPG (Germany) and IGN (Spain). The CSO is operated by the California Institute of Technology under cooperative agreement with the National Science Foundation (AST-0838261).
\bibliographystyle{mn2e}
\bibliography{biblio_G34}

\appendix

\newpage

\section{\textit{Herschel}/HIFI observations}

 The observing IDs of the \textit{Herschel}/HIFI data are listed in Table \ref{obsid}. 

\begin{table*}
\renewcommand{\arraystretch}{1.2}
\caption{List of the \textit{Herschel}/HIFI obsIDs}
\begin{tabular}{l c c c c c c }
\hline \hline
Species & Frequency (GHz) & Transition  &  ObsID-A  & ObsID-B & ObsID-C & Observing program \\
\hline
HDO & 490.5966 & 2$_{0,2}$--1$_{1,1}$ & 1342244320 & 1342244321 & 1342244322 & OT1 \\    
HDO & 509.2924 & 1$_{1,0}$--1$_{0,1}$ & 1342219186 & 1342219187 & 1342219188 & PRISMAS \\    
HDO & 599.9267 & 2$_{1,1}$--2$_{0,2}$ & 1342230375 & 1342230376 & 1342230377 & OT1  \\     
HDO & 848.9618 & 2$_{1,2}$--1$_{1,1}$ &  1342244114 & 1342244115 & 1342244116 & OT1 \\  
HDO & 893.6387 & 1$_{1,1}$--0$_{0,0}$ &  1342207360 & 1342207361 & 1342207362 & PRISMAS \\  
HDO & 919.3109 & 2$_{0,2}$--1$_{0,1}$ & 1342244381 & 1342244382 & 1342244383 & OT1  \\  
ortho--H$_2^{18}$O & 547.6764 & 1$_{1,0}$--1$_{0,1}$ & 1342194468 &1342194469 & 1342194470 & PRISMAS \\
para--H$_2^{18}$O & 1101.6983 & 1$_{1,1}$--0$_{0,0}$ & 1342207367 & 1342207368 &1342207369 & PRISMAS \\
\hline \\
\end{tabular}
\label{obsid}
\end{table*}%


\section{Non-LTE spherical radiative transfer modeling}

\subsection{Density and temperature profiles}

\begin{figure}
\begin{center}
\includegraphics[width=0.45\textwidth]{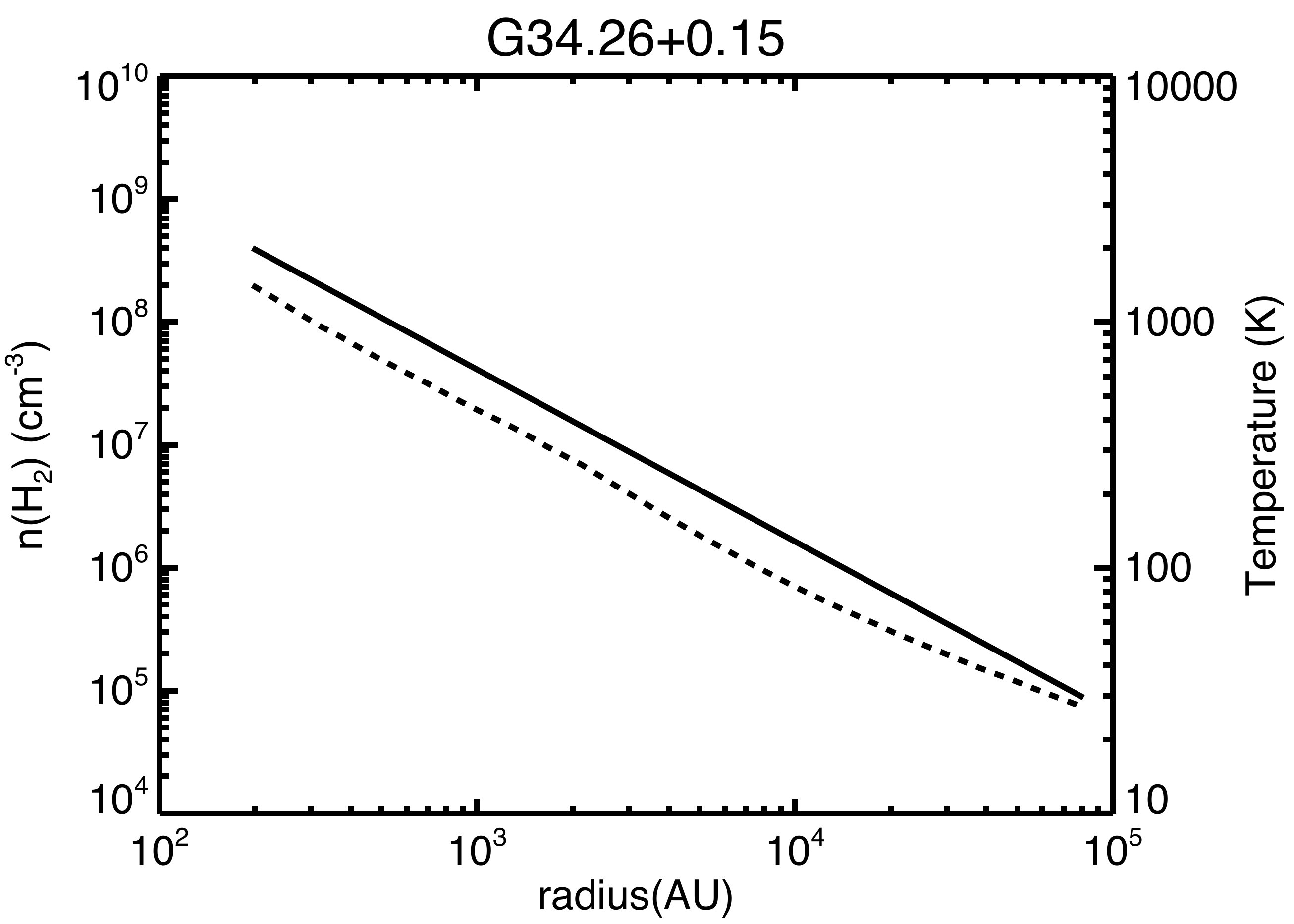}
\caption{H$_2$ density (solid line) and temperature (dashed line) profiles from \citet{vanderTak2013}.} 
\label{phys_structure}
\end{center}
\end{figure}

The density and temperature profiles used for the model of the HDO and H$_2^{18}$O lines are shown in Figure \ref{phys_structure}.

\subsection{Velocity field profiles}
\label{app_vel}

The radial velocity profile ($\varv_r$) and the turbulence width (Doppler b-parameter, $db$) are required inputs for the RATRAN radiative transfer modeling. We do not have direct information on them, except that inward motions are necessary to reproduce the inverse P-Cygni profile observed on the HDO 1$_{1,1}$--1$_{0,0}$ fundamental line observed with HIFI. Indeed, with a static envelope, the model predicts an absorbing component at the local standard velocity of rest, $V_{\rm LSR}$ = 58 \kms, whereas the absorption is observed at 61 \kms~only. Consequently, it is not possible to reproduce the line profile of this HDO fundamental transition with a static envelope and an unrelated absorbing layer situated on the line of sight. The radial velocity was then fixed at -3 km\,s$^{-1}$ in the cold outer regions. 

To estimate the $\varv_r$ and $db$ profiles throughout the envelope, we assumed abundance profiles with a jump and proceeded as follows: \textit{i)} First, we assume an initial profile for the radial velocity and the Doppler b-parameter. Then we run a grid of models with various inner ($T > T_{\rm j}$) and outer ($T < T_{\rm j}$) abundances. \textit{ii)} If the predicted line profiles do not fit at all the data, we choose the model that gives the best agreement with respect to the intensities. Keeping the same abundances as this model, we modify the velocity profiles to obtain better agreement. \textit{iii)} Then we run another grid of models with these new velocity profiles and go back to step \textit{(ii)} if necessary and so on.

Models with infall profiles were attempted but, as shown in Figure \ref{ratran_hdo_outward_vs_inward}, some of the HDO lines (509, 600, 849, and 919 GHz) are shifted in velocity with respect to the observations when this type of profile is taken into account.
On the contrary, models with outward motions in the inner regions give a good agreement for the different HDO lines. These outward motions could be produced by stellar winds or outflows.

\begin{figure}
\begin{center}
\includegraphics[width=0.45\textwidth]{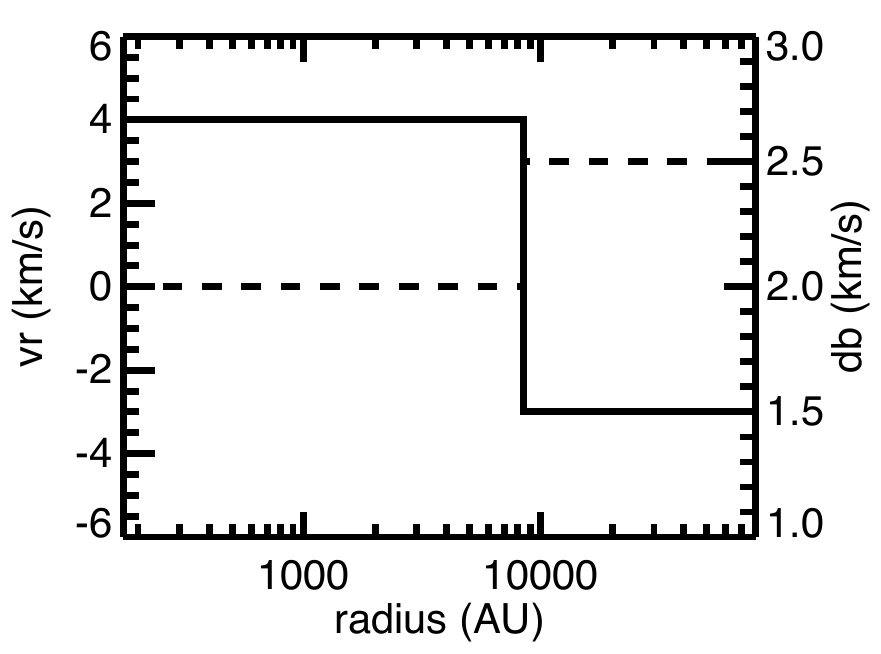} 
\caption{Radial velocity ($\varv r$, solid line) and Doppler b-parameter  ($db$, dashed line) profiles assumed for the different models presented in the paper.}
\label{vel_prof}
\end{center}
\end{figure}

\begin{figure*}
\begin{center}
\includegraphics[width=0.9\textwidth]{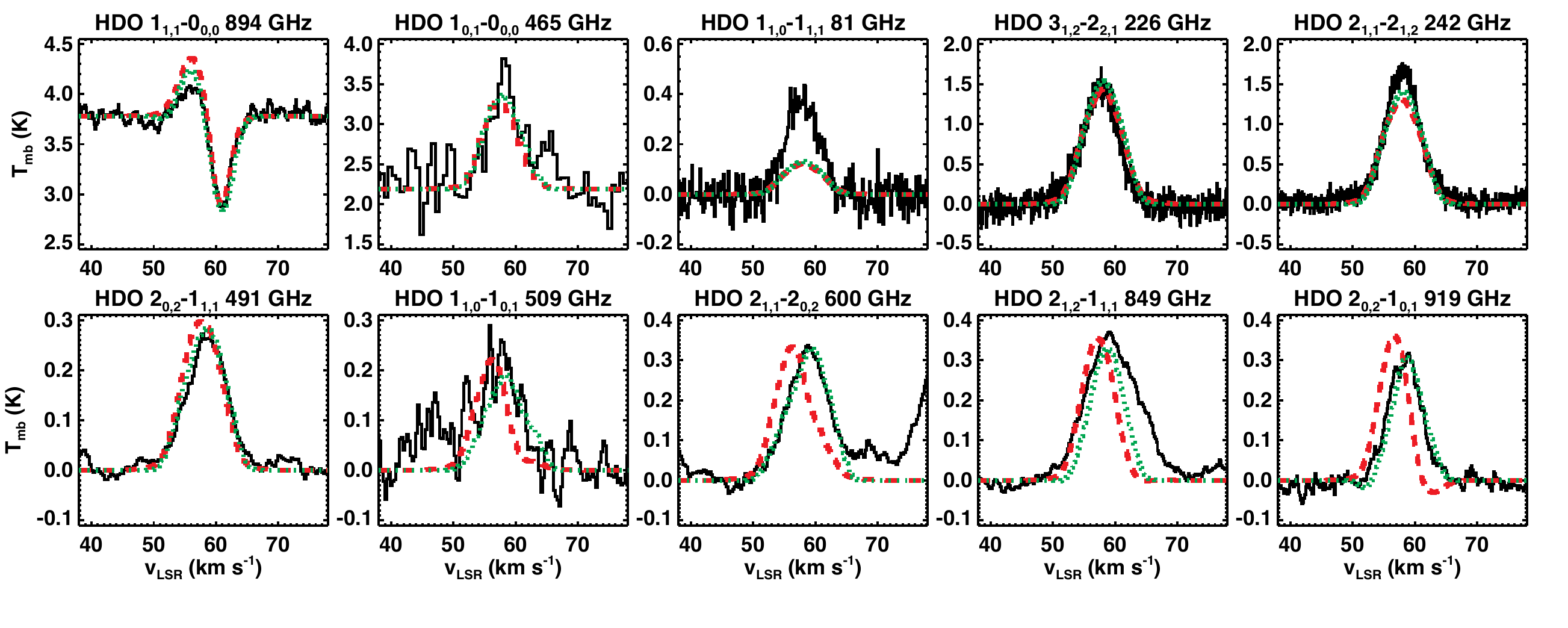}
\caption{{\it Black solid line}: HDO lines observed with HIFI, IRAM, and CSO.
{\it Red dashed line}: Modeling with an infall profile (free-fall model with $M$ = 20 $M_\odot$). {\it Green dotted line}: Modeling with a velocity profile with inward motions in the outer regions and outward motions in the inner regions. }
\label{ratran_hdo_outward_vs_inward}
\end{center}
\end{figure*}

To limit the number of free parameters in the study, we only considered constant $\varv_r$ and $db$ values in the inner and outer regions.
Based on the widths of the different lines, the $\varv_r$ parameter was determined to be about 4\,km\,s$^{-1}$ in the inner regions in expansion and about -3\,km\,s$^{-1}$ in the infalling outer regions. The Doppler b-parameter is estimated to be about 2.5\,km\,s$^{-1}$ in the outer region. This constraint is based on the width of the absorption component of the HDO line at 894 GHz.  In the inner regions, the fit of the emission lines is slightly better with $db$ $\sim$ 2.0\,km\,s$^{-1}$. The $db$ parameter seems to decrease from the outer to the inner regions, similarly to other studies in high mass sources \citep{Caselli1995,Herpin2012}.
To be able to reproduce the HDO and H$_2^{18}$O lines with a unique velocity profile for all the models with different jump temperatures , we delimited the change of the $db$ and $\varv_r$ values between the inner and outer region at 100\,K.
The final velocity fields used in the paper are presented in Figure \ref{vel_prof}. We cannot however exclude that different velocity fields (possibly more complex) would reproduce the lines.
For the models assuming an increase of the abundance between 100 and 200\,K (see Section \ref{sect_mod_increase}), we also checked that a radial velocity increasing gradually from -3 to 4 km\,s$^{-1}$ between 100 and 200\,K would give similar results.

\subsection{Modeling of the HDO lines with a two-jump modeling }
\label{app_2jumps}

In their study of Sgr B2(M), \citet{Comito2010} showed that a model with two abundance jumps, one at 100\,K and another at 200\,K, is necessary to reproduce the different HDO lines observed towards this star-forming region.
To check the influence on our results, we ran grids with such assumptions for our modeling of G34. Several models give a good agreement with the data. However the fit to the data is not better than with one jump. It is extremely similar to the model shown in Figure \ref{ratran_hdo_200K}. Unsurprisingly, the best-fit gives an inner abundance ($T$ $>$ 200\,K) of 2\,$\times$\,10$^{-7}$, consistent with a modeling with a unique jump at a temperature of 200\,K.
The outer abundance ($T$ $<$ 100\,K) is estimated to be 8\,$\times$\,10$^{-11}$, and the abundance between 100 and 200\,K is only constrained by an upper limit of 1\,$\times$\,10$^{-9}$. 
Indeed, with abundances higher than 1\,$\times$\,10$^{-9}$, the predicted intensities for the lines at 491, 600 and 919 GHz become too high compared with the observations.
These results are consistent with a modeling having a single jump, and the two-jump assumption does not improve the fit of the model results to the data.
Although we cannot exclude that a double abundance jump occurs here, we estimate that the hypothesis of one jump is probably more reasonable. Indeed, chemical reasons for a three-step abundance profile are unclear. All the HDO trapped in the grain mantles should desorb thermally at approximately 100\,K. Additional formation of water is possible into the gas phase at higher temperatures. But due to the relatively high temperatures in the hot core, the formation of deuterated water should be negligible. In addition, the HDO abundances derived between 100 and 200\,K are very low ($\leq$ 1\,$\times$\,10$^{-9}$) comparatively to the abundance above 200\,K ($\sim$\,2\,$\times$\,10$^{-7}$).
Uncertainty in the temperature profile in the inner region with a single jump produced by the desorption from grain mantles or a gradual increase of the abundance probably provide a better explanation than a two jump model. 

\subsection{Summary of the different models}

\begin{table*}
\caption{List of the models presented in the paper.}
\begin{center}
\begin{tabular}{c l l }
\hline \hline
Model & Abundance profile & Sections \& Figures \\
\hline
 1 & 1 jump $T_{\rm j}$ = 100 K & HDO : Sect. \ref{sect_jump}, Fig. \ref{ratran_hdo_100K} \\
 \hline
 2 & 1 jump $T_{\rm j}$ = 120 K & HDO : Sect. \ref{sect_jump}, Fig. \ref{ratran_hdo_120K}\\
  \hline
 3 & 1 jump $T_{\rm j}$ = 150 K & HDO : Sect. \ref{sect_jump}, Fig. \ref{ratran_hdo_150K}\\
  & &  H$_2^{18}$O : Sect. \ref{sect_h218o}, Fig. \ref{ratran_h218o_150K} \\
   \hline
 4 & 1 jump $T_{\rm j}$ = 180 K & HDO : Sect. \ref{sect_jump}, Fig. \ref{ratran_hdo_180K}  \\
  & &  H$_2^{18}$O : Sect. \ref{sect_h218o}, Fig. \ref{ratran_h218o_180K} \\
   \hline
  5 & 1 jump $T_{\rm j}$ = 200 K & HDO : Sect. \ref{sect_jump}, Fig. \ref{ratran_hdo_200K}  \\
  & &  H$_2^{18}$O : Sect. \ref{sect_h218o}, Fig. \ref{ratran_h218o_200K} \\
   \hline
 5 & 1 jump $T_{\rm j}$ = 220 K & HDO : Sect. \ref{sect_jump}, Fig. \ref{ratran_hdo_220K} \\
  & &  H$_2^{18}$O : Sect. \ref{sect_h218o}, Fig. \ref{ratran_h218o_220K} \\
  \hline
  6 & 2 jumps $T_{\rm j1}$ = 100 K, $T_{\rm j2}$ = 200 K & HDO : Appx. \ref{app_2jumps} \\
\hline
7 & Constant inner abundance ($T$ $\geq$ 200\,K) and   & HDO : Sect. \ref{sect_slope}, Fig. \ref{ratran_hdo_Gcr}  \\
 &  decrease of the outer abundance from the & H$_2^{18}$O : Sect. \ref{sect_slope}, Fig. \ref{ratran_h218o_Gcr} \\
 & cold to the warm regions (see Fig. \ref{water_profiles}) &  \\
 \hline
 8 & Constant inner abundance, gradual  & HDO : Sect. \ref{sect_mod_increase}, Fig. \ref{ratran_hdo_Gcr}  \\
 &  increase of the abundance at the transition & H$_2^{18}$O : Sect. \ref{sect_mod_increase}, Fig. \ref{ratran_h218o_Gcr}\\
 &  hot core/cold envelope (100--200\,K) and decrease & \\
&   of the outer abundance from the cold regions  \\
& to the regions at 100\,K (see Fig. \ref{water_profiles}) \\
 \hline
\end{tabular}
\end{center}
\label{table_summary_run_models}
\end{table*}%

 Table \ref{table_summary_run_models} summarizes the different types of models shown in the paper.

\begin{figure*}
\begin{center}
\includegraphics[width=0.9\textwidth]{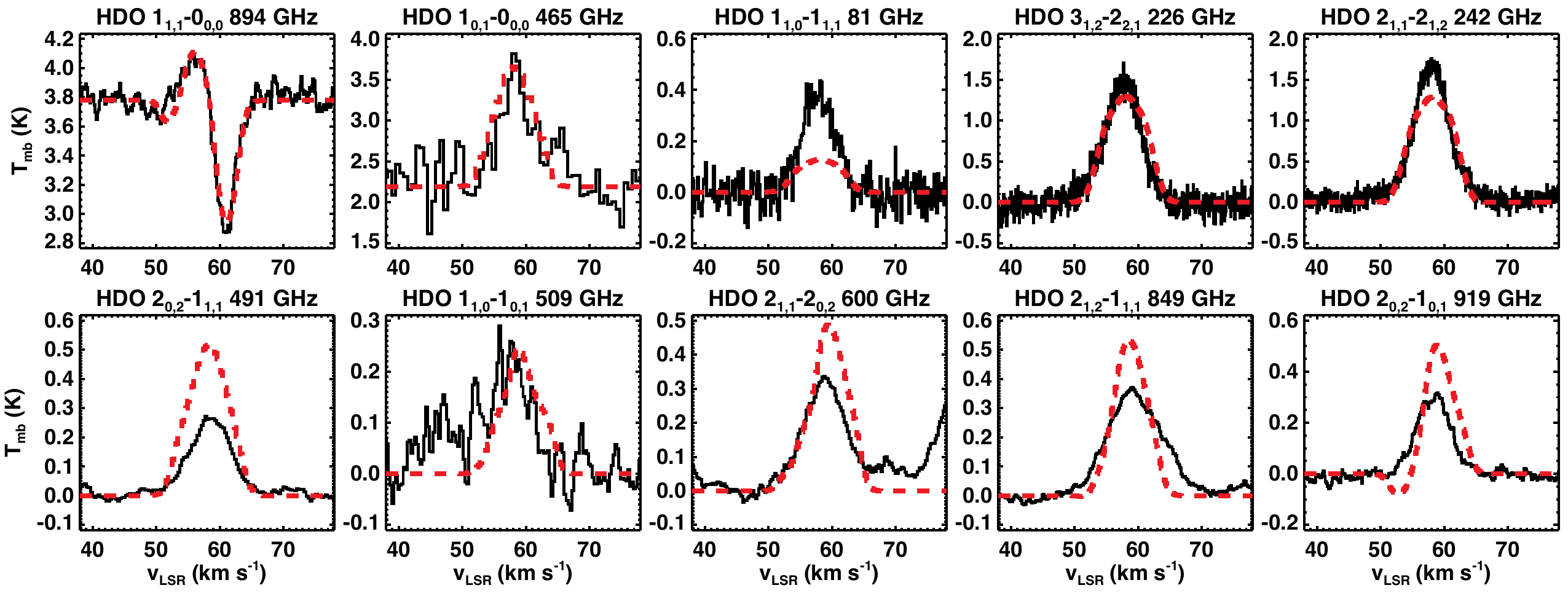}
\caption{{\it Black solid line}: HDO lines observed with HIFI, IRAM, and CSO.
{\it Red dashed line}: Modeling for a jump temperature $T_{\rm j}$ = 120 K, an inner abundance $X_{\rm in}$ = 6\,$\times$\,10$^{-8}$ and an outer abundance $X_{\rm out}$ = 8\,$\times$\,10$^{-11}$. }
\label{ratran_hdo_120K}
\end{center}
\end{figure*}

\begin{figure*}
\begin{center}
\includegraphics[width=0.9\textwidth]{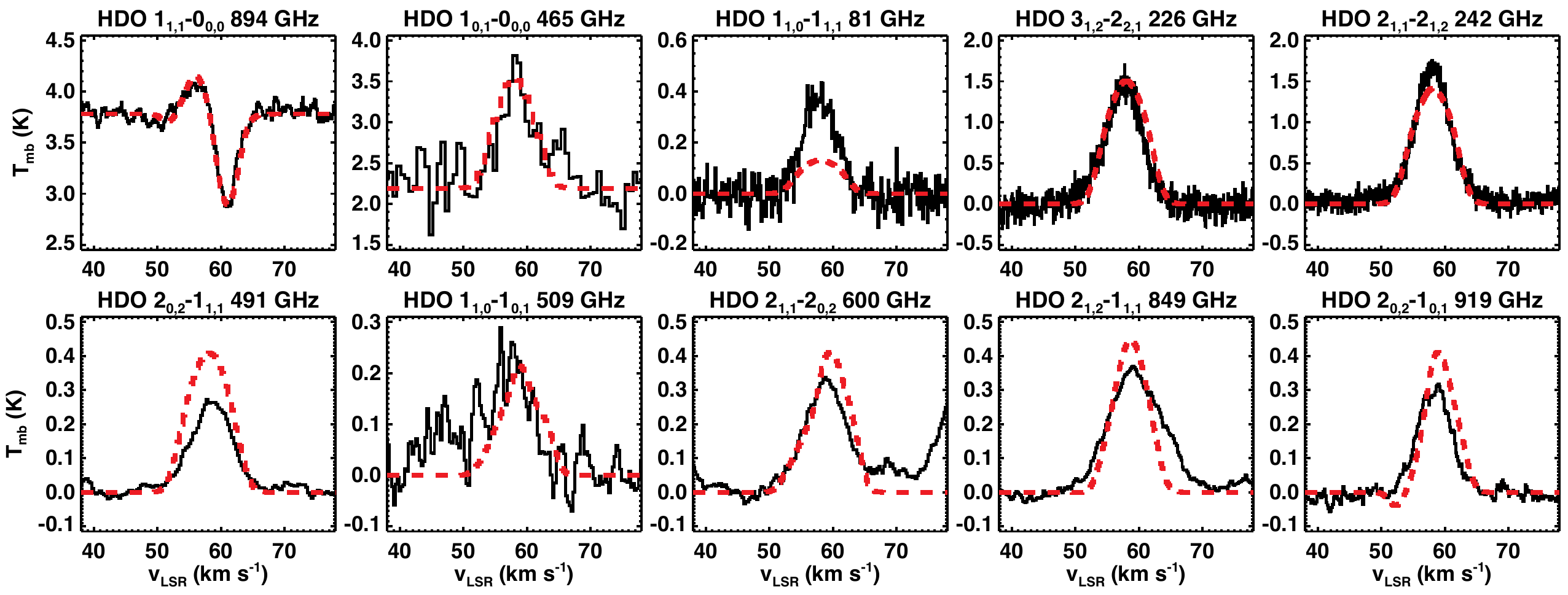}
\caption{{\it Black solid line}: HDO lines observed with HIFI, IRAM, and CSO.
{\it Red dashed line}: Modeling for a jump temperature $T_{\rm j}$ = 150 K, an inner abundance $X_{\rm in}$ = 1\,$\times$\,10$^{-7}$ and an outer abundance $X_{\rm out}$ = 8\,$\times$\,10$^{-11}$. }
\label{ratran_hdo_150K}
\end{center}
\end{figure*}

\begin{figure*}
\begin{center}
\includegraphics[width=0.9\textwidth]{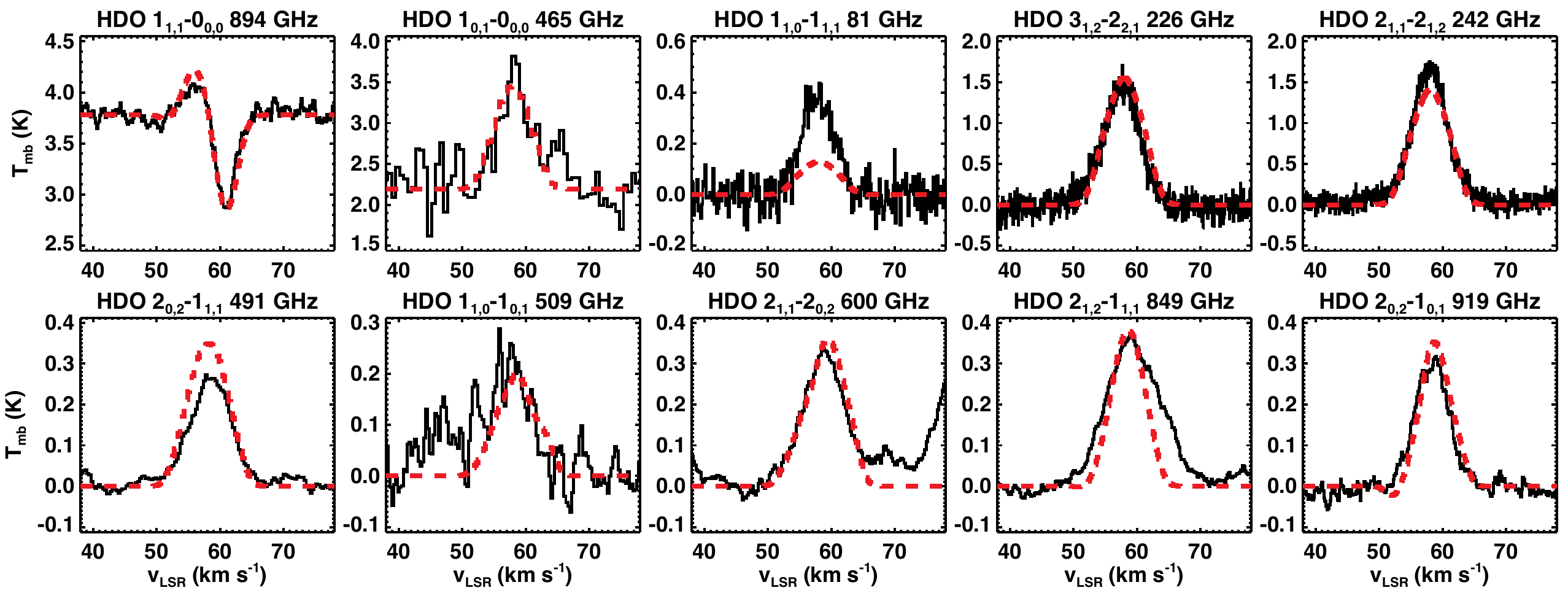}
\caption{{\it Black solid line}: HDO lines observed with HIFI, IRAM, and CSO.
{\it Red dashed line}: Modeling for a jump temperature $T_{\rm j}$ = 180 K, an inner abundance $X_{\rm in}$ = 1.5\,$\times$\,10$^{-7}$ and an outer abundance $X_{\rm out}$ = 8\,$\times$\,10$^{-11}$. }
\label{ratran_hdo_180K}
\end{center}
\end{figure*}

\begin{figure*}
\begin{center}
\includegraphics[width=0.9\textwidth]{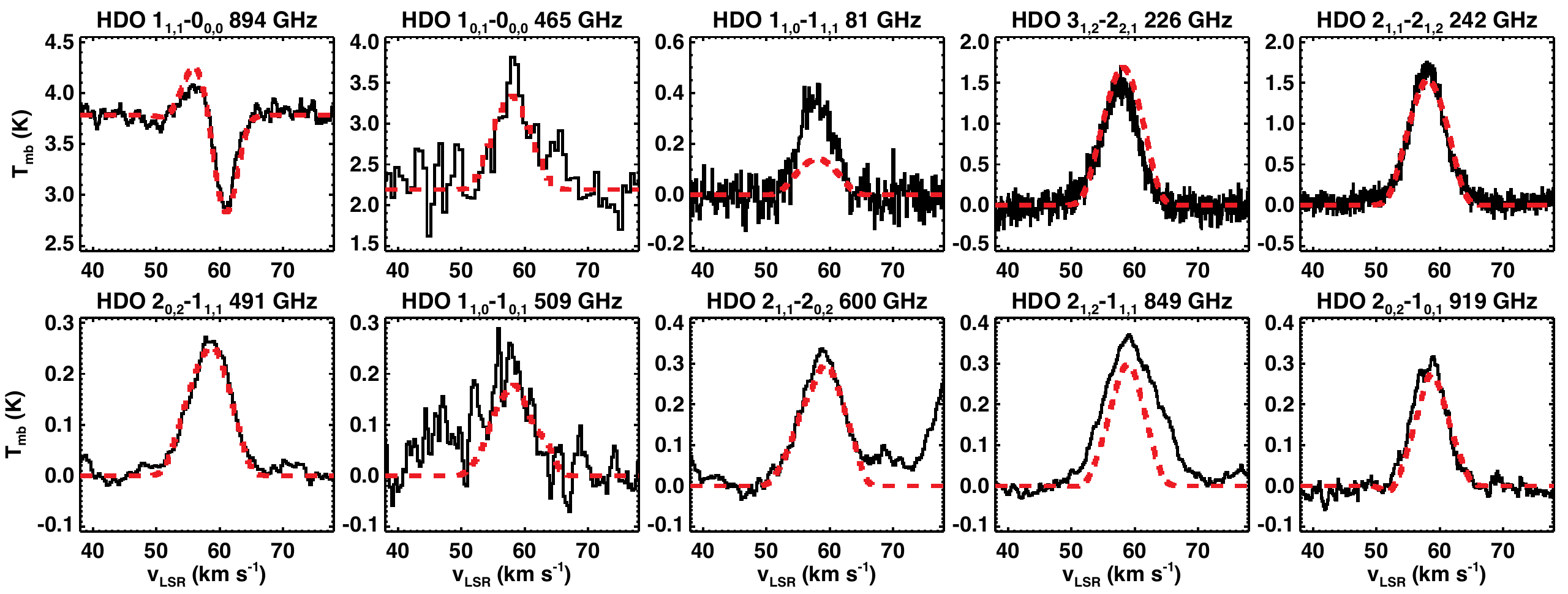}
\caption{{\it Black solid line}: HDO lines observed with HIFI, IRAM, and CSO.
{\it Red dashed line}: Modeling for a jump temperature $T_{\rm j}$ = 220 K, an inner abundance $X_{\rm in}$ = 3\,$\times$\,10$^{-7}$ and an outer abundance $X_{\rm out}$ = 8\,$\times$\,10$^{-11}$. }
\label{ratran_hdo_220K}
\end{center}
\end{figure*}

\begin{figure*}
\begin{center}
\includegraphics[width=0.62\textwidth]{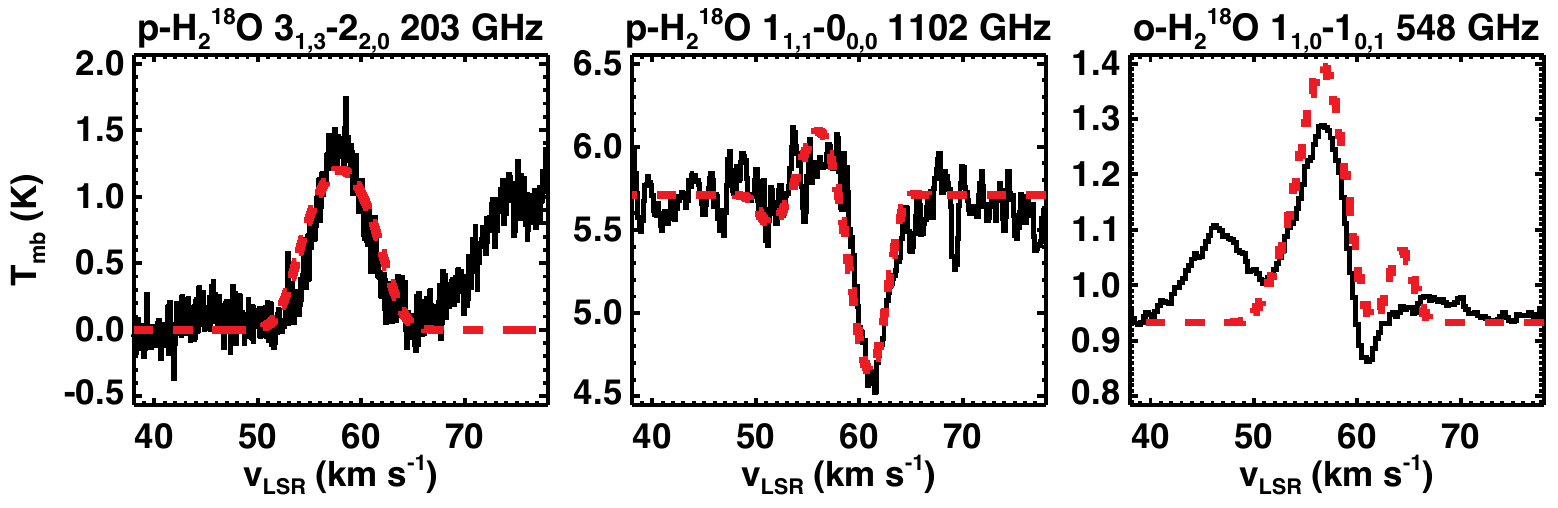}
\caption{{\it Black solid line}: H$_2^{18}$O lines observed with HIFI and IRAM.
{\it Red dashed line}: Modeling for a jump temperature $T_{\rm j}$ = 150 K, an inner abundance $X_{\rm in}$ = 4\,$\times$\,10$^{-7}$ and an outer abundance $X_{\rm out}$ = 1.3\,$\times$\,10$^{-10}$. }
\label{ratran_h218o_150K}
\end{center}
\end{figure*}

\begin{figure*}
\begin{center}
\includegraphics[width=0.62\textwidth]{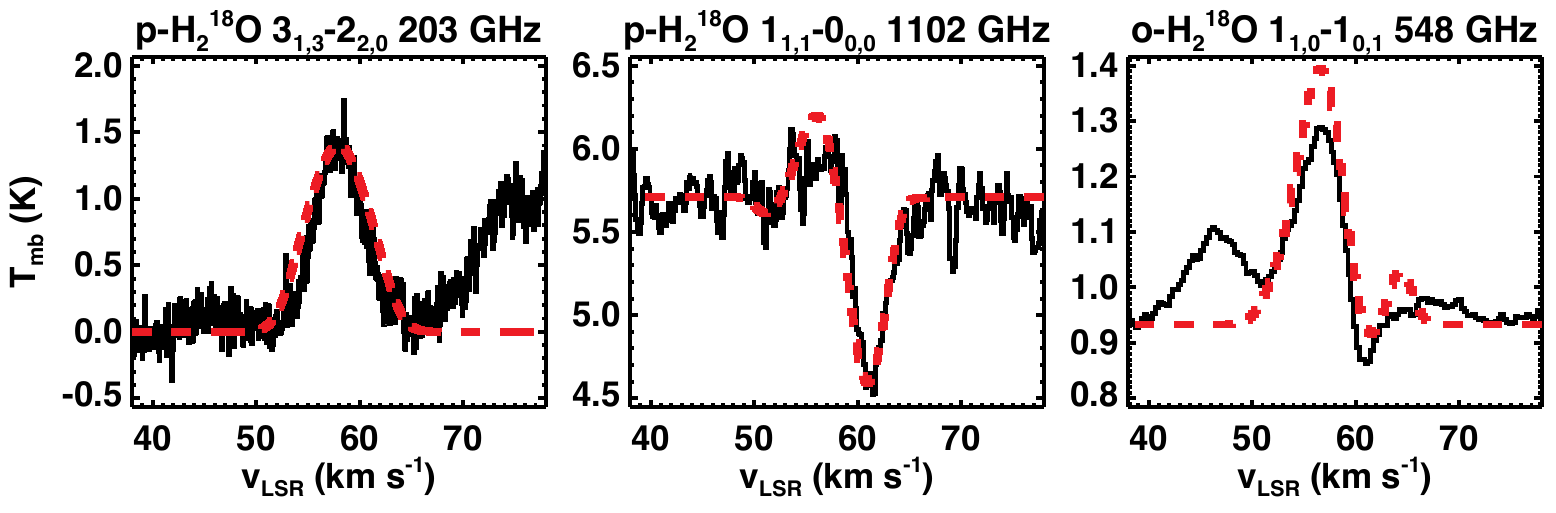}
\caption{{\it Black solid line}: H$_2^{18}$O lines observed with HIFI and IRAM.
{\it Red dashed line}: Modeling for a jump temperature $T_{\rm j}$ = 180 K, an inner abundance $X_{\rm in}$ = 7\,$\times$\,10$^{-7}$ and an outer abundance $X_{\rm out}$ = 1.3\,$\times$\,10$^{-10}$. }
\label{ratran_h218o_180K}
\end{center}
\end{figure*}

\begin{figure*}
\begin{center}
\includegraphics[width=0.62\textwidth]{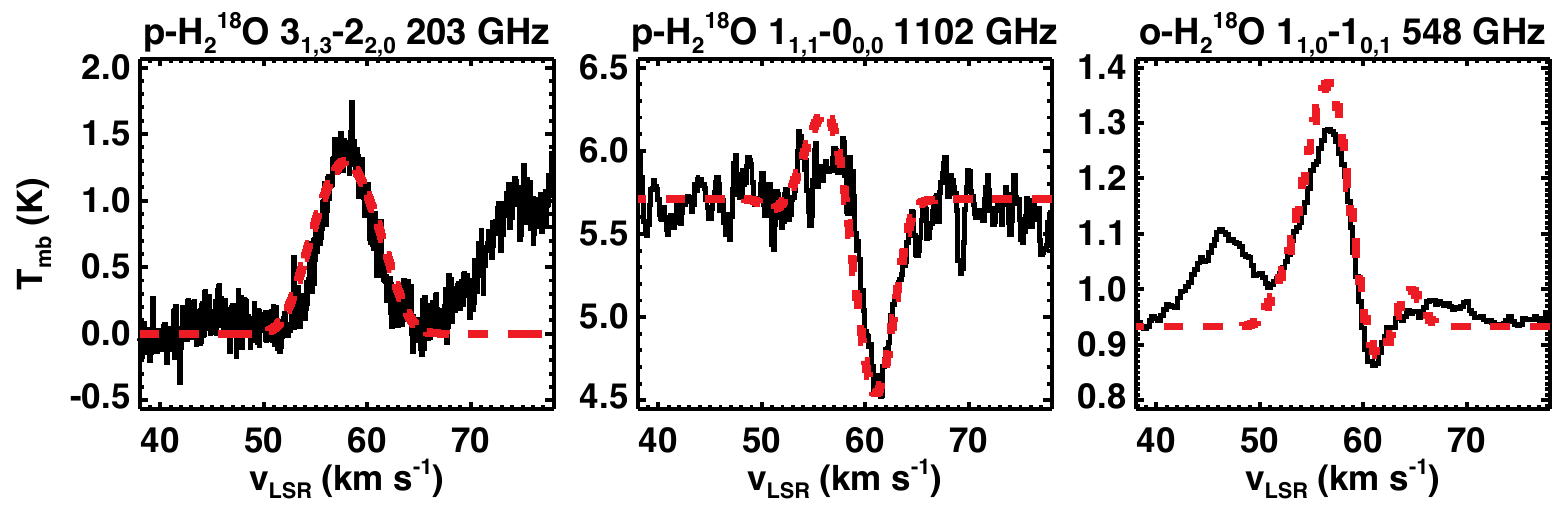}
\caption{{\it Black solid line}: H$_2^{18}$O lines observed with HIFI and IRAM.
{\it Red dashed line}: Modeling for a jump temperature $T_{\rm j}$ = 220 K, an inner abundance $X_{\rm in}$ = 1.2\,$\times$\,10$^{-6}$ and an outer abundance $X_{\rm out}$ = 1.3\,$\times$\,10$^{-10}$. }
\label{ratran_h218o_220K}
\end{center}
\end{figure*}

\section{Sensitivity of our modeling to the inclusion of spin-state chemistry}
\label{apdx:without_opm}

\begin{figure}
\begin{center}
\includegraphics[width=1.0\linewidth]{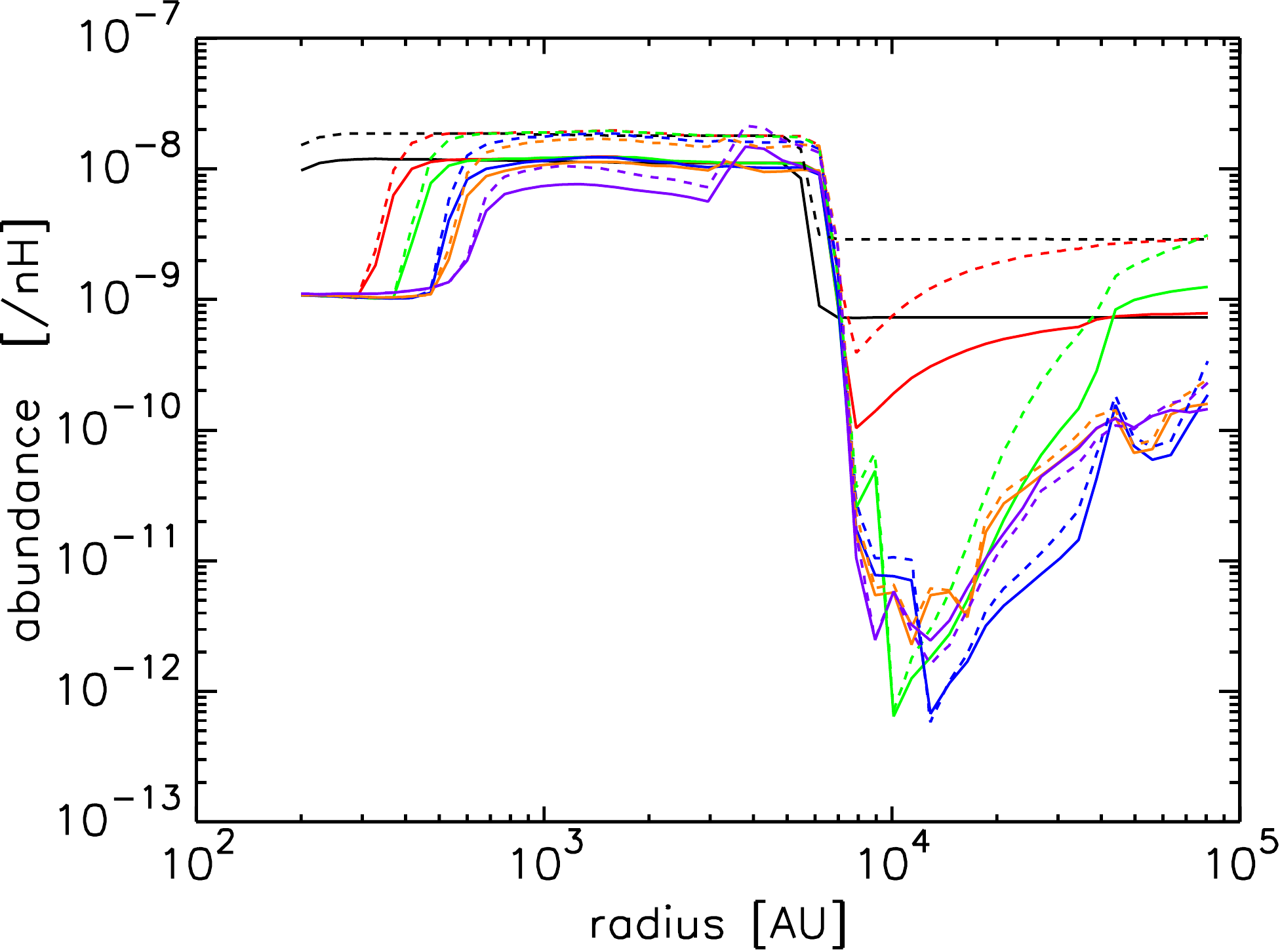}
\includegraphics[width=1.0\linewidth]{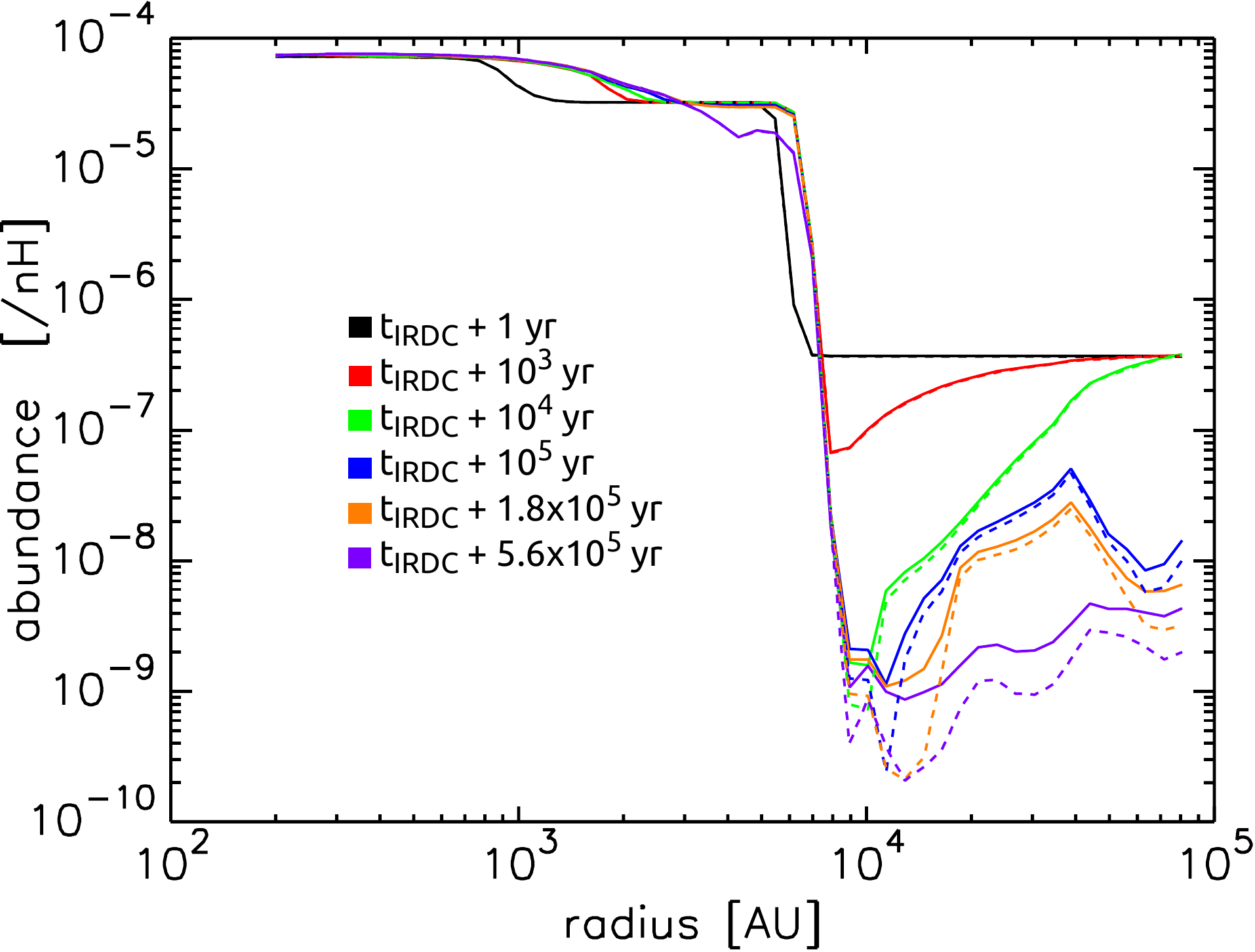}
\includegraphics[width=1.0\linewidth]{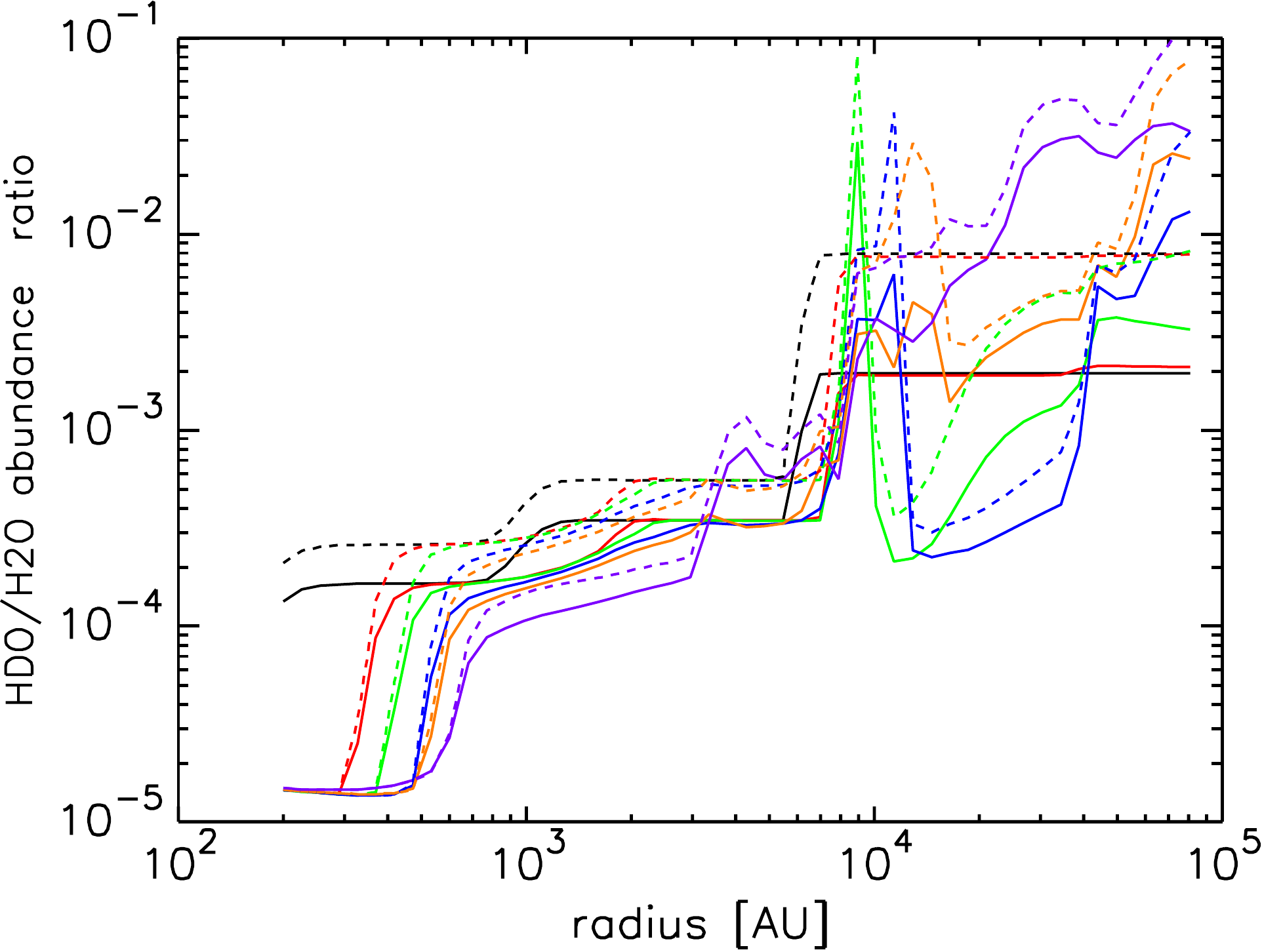}
\caption{Calculated gas-phase abundances of HDO (top panel) and H$_2$O (middle panel) vs radius with (solid lines) and without (dashed lines) spin-state chemistry.
The bottom panel shows the corresponding HDO/H$_2$O ratios.
The results are time dependent, and the colors correspond to different values after the first initial phase as in figure~\ref{fig_chemistry_HDO_H2O}.}
\label{fig:HDO_H2O_ratio_test-opm}
\end{center}
\end{figure}

We studied the sensitivity of our results to the chemistry involving the spin states of H$_2$, D$_2$, H$_3^+$, H$_2$D$^+$, D$_2$H$^+$, and D$_3^+$.
Using the same physical condition as described in section~\ref{subsubsec:model}, we modeled the chemical evolution of our source using our basic chemical network adapted from \citet{Aikawa2012} and \citet{Furuya2012}, without taking into account spin states.
Figure~\ref{fig:HDO_H2O_ratio_test-opm} presents calculated gaseous HDO and H$_2$O abundances, and HDO/H$_2$O ratios, with and without spin-state chemistry.

The gaseous HDO abundance is higher throughout the source by a factor 2 (inner core) to 4 (cold envelope), if we neglect spin state chemistry.
We expected this result since the inclusion of spin states tends to reduce deuterium fractionation by populating ortho spin states of H$_{2}$.
The effect is more pronounced in the cold envelope because the endothermic reactions involving deuterated ions and o-H$_{2}$ are greatly enhanced in rate.  
The differences are time dependent, and are less pronounced as the system progresses from the IRDC phase to the end of the simulation because o-H$_2$ is converted in p-H$_2$.

In the hot inner region ($\leq10^4$~AU) both models give the same results for gas-phase water.
In the cold envelope, there is more gas phase water if we take into account spin-state chemistry.
Contrary to HDO, the differences grow larger with time, and this trend is particularly strong in the transition area ($\simeq10^4$~AU).
In this area, the temperature is low enough to not allow efficient desorption, and the density is high so adsorption is important. In consequence, water is primarily on the grain surfaces.
In this situation, gas phase reactions are the principal pathway to form gas phase water, mainly through H$_3$O$^+$ : the spin-state chemistry reduces the formation of H$_2$D$^+$ from H$_3^+$, and the extra amount of H$_3^+$ helps to produce more H$_3$O$^+$ through successive reactions involving OH, H$_2$O$^+$, and H$_2$.

Globally, since the HDO abundance is lower and the H$_2$O abundance is higher if we take into account spin-state chemistry, the HDO/H$_2$O ratio is lower.
The decrease is within a factor of 10, and depends on the radius and the time.
The differences are less critical at high temperature than at low temperature, but are not negligible compared with the variation of our measured ratios.

\twocolumn

\end{document}